\documentclass[prd,preprint,superscriptaddress,nofootinbib,longbibliography]{revtex4-1}
\usepackage{graphicx}
\usepackage{amsmath}
\usepackage{amsfonts}
\usepackage{mathtools}
\usepackage{color}
\usepackage{amssymb}
\usepackage{dcolumn}
\usepackage{bm}
\usepackage{braket}
\usepackage{microtype} 
\usepackage[linktoc=all]{hyperref}
\usepackage{cleveref}

\numberwithin{equation}{section}
\newcommand{\lag}{\mathcal{L}}
\newcommand{\ud}{\mathrm{d}}

\newcommand{\bel}[1] {\begin{equation}\label{#1}}
\newcommand{\beal}[1] {\begin{eqnarray}\label{#1}}

\newcommand{\be}{\begin{equation}}
\newcommand{\bea}{\begin{eqnarray}}
\newcommand{\ee}{\end{equation}}
\newcommand{\eea}{\end{eqnarray}}

\makeatletter
\DeclareRobustCommand{\rcite}[1]{%
\rcite@aux#1,\@nil{#1}%
}
\def\rcite@aux#1,#2\@nil#3{%
\if\relax#2\relax
Ref.~\cite{#3}%
\else
Refs.~\cite{#3}%
\fi
}
\makeatother

\begin{document}
\title{Radiation of scalar modes and the classical double copy}

\author{Mariana Carrillo Gonz\'alez} 
\email{cmariana@sas.upenn.edu}
\affiliation{Center for Particle Cosmology, Department of Physics and Astronomy,
	University of Pennsylvania, Philadelphia, Pennsylvania 19104, USA}
\author{Riccardo Penco}
\email{rpenco@sas.upenn.edu}
\affiliation{Center for Particle Cosmology, Department of Physics and Astronomy,
	University of Pennsylvania, Philadelphia, Pennsylvania 19104, USA}
\author{Mark Trodden}
\email{trodden@upenn.edu}
\affiliation{Center for Particle Cosmology, Department of Physics and Astronomy,
	University of Pennsylvania, Philadelphia, Pennsylvania 19104, USA}

\date{\today}

\begin{abstract}
\noindent 
The double copy procedure relates gauge and gravity theories through color-kinematics replacements, and holds for both scattering amplitudes and in classical contexts. Moreover, it has been shown that there is a web of theories whose scattering amplitudes are related through operations that exchange color and kinematic factors. In this paper, we generalize and extend this procedure by showing that the classical perturbative double copy of pions corresponds to special Galileons. We consider point-particles coupled to the relevant scalar fields, and find the leading and next to leading order radiation amplitudes. By considering couplings motivated by those that would arise from extracting the longitudinal modes of the gauge and gravity theories, we are able to map the non-linear sigma model radiation to that of the special Galileon. We also construct the single copy by mapping the bi-adjoint scalar radiation to the non-linear sigma model radiation through generalized color-kinematics replacements.
\end{abstract}

\maketitle

\tableofcontents

\newpage

\section{Introduction}
A surprising relationship between gauge theories and gravity has been shown to exist not only for scattering amplitudes but also in more general cases. In the scattering amplitudes context, an example of this relationship is the BCJ (Bern, Carrasco, Johannson) double copy \cite{Bern:2008qj,Bern:2010yg,Bern:2010ue} which consists of applying color-kinematics replacements to the Yang-Mills scattering amplitudes in order to obtain the scattering amplitudes of a gravitational theory involving a graviton, a dilaton, and a two-form field. In this case, there is a duality between the color factors and the kinematic factors, since both can satisfy the same algebra. In the cases where the double copy maps observables other than scattering amplitudes, the idea of performing color-kinematics replacements persists, but the existence of an algebra satisfied by the analogue of the kinematic factors has been scarcely explored. An example of these duality satisfying kinematic factors in the classical context was introduced in \cite{Shen:2018ebu}.  Some of the new cases consist of a classical realization of the double copy and follow two main directions: exact results \cite{Monteiro:2014cda,Luna:2015paa,Luna:2016due,Ridgway:2015fdl,Carrillo-Gonzalez:2017iyj,Bahjat-Abbas:2017htu,Lee:2018gxc,Ilderton:2018lsf,Berman:2018hwd}, and perturbative results \cite{Saotome:2012vy,Neill:2013wsa,Luna:2016hge,Goldberger:2016iau,Goldberger:2017frp,Goldberger:2017vcg,Goldberger:2017ogt,Chester:2017vcz,Shen:2018ebu,Plefka:2018dpa}. While the  construction of the double copy in the case of amplitudes relies heavily on the cubic structure (or rather the ability to write the theory in a form in which there is a cubic interaction), in the classical double copy this technical requirement is not always obvious. In the case of exact results, one starts from the gravitational side with a solution in the form of a Kerr-Schild metric and applies the corresponding color-kinematics replacements to obtain the single copy, {\it i.e..}, the gauge theory analogue. Applying these replacements one more time leads to the bi-adjoint scalar analogue, the zeroth copy. In the perturbative case, one can take the opposite direction and start from the bi-adjoint scalar, then apply the corresponding replacements and obtain Yang-Mills theory, perform this one more time and obtain the gravitational theory. Other surprising examples of the double copy have been discovered in different contexts---see for example \cite{Anastasiou:2014qba,Anastasiou:2017nsz,Borsten:2015pla,Cardoso:2016amd,Cardoso:2016ngt,Chu:2016ngc,Mizera:2018jbh}. In this paper, we will focus on the perturbative implementation of the classical double copy.

\begin{figure}[!t]
	\includegraphics[scale=1.4]{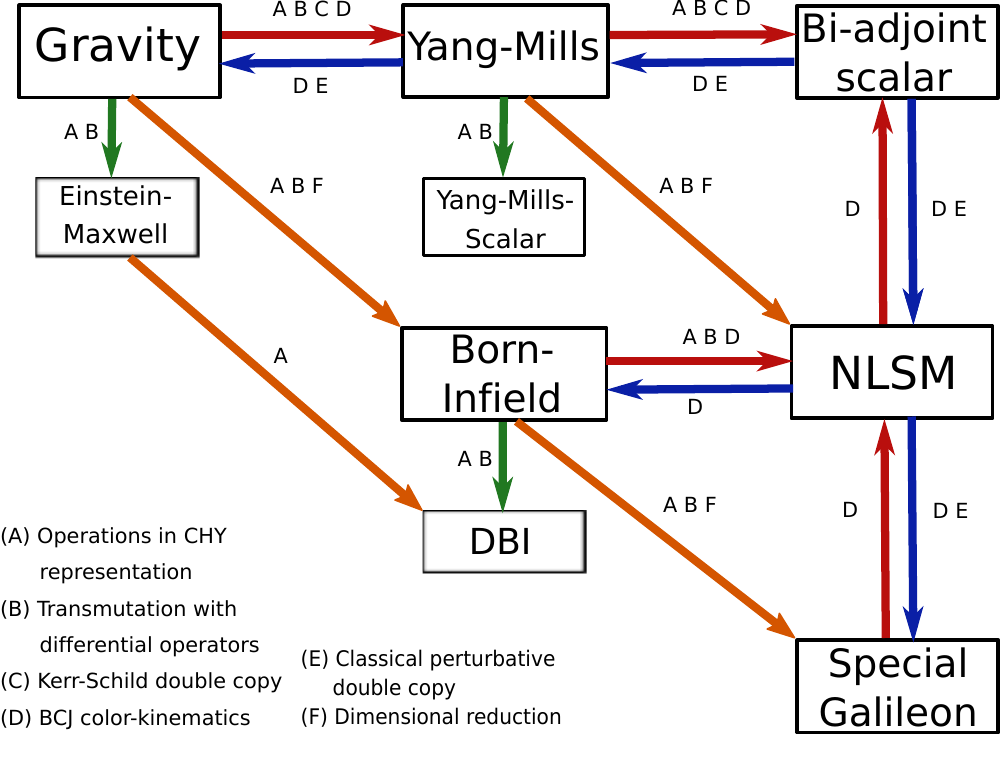}
	\caption{Web of relationships between various field theories. 
		The operations in CHY representation correspond to those in \cite{Cachazo:2014xea} and the transmutations with differential operators are given in \cite{Cheung:2017ems}. Some examples of the Kerr-Schild double copy can be found in \cite{Monteiro:2014cda,Luna:2015paa,Luna:2016due,Ridgway:2015fdl,Carrillo-Gonzalez:2017iyj,Bahjat-Abbas:2017htu,Lee:2018gxc,Ilderton:2018lsf,Berman:2018hwd}. For examples of the BCJ double copy, see \cite{Elvang:2015rqa,Cheung:2017pzi,DoubleCopyLectureNotes} and references there in.  Classical perturbative double copy examples are found in \cite{Goldberger:2016iau,Goldberger:2017frp,Goldberger:2017ogt,Goldberger:2017vcg,Chester:2017vcz,Shen:2018ebu}. The dimensional reduction refers to that in \cite{Cheung:2017yef}. Besides the relations shown in this figure, there are other cases of relations between extended theories; some of these examples are found in \cite{Carrasco:2016ygv,Chiodaroli:2017ngp,Nandan:2016pya}. } 
	\label{fig:web}
\end{figure}

Such a procedure applies more broadly than between gauge and gravitational theories. For example, by considering a ``dimensional reduction" of the gauge and gravity theories one can obtain the scattering amplitudes of the non-linear sigma model (NLSM) and the special Galileon, respectively \cite{Cachazo:2014xea,Cheung:2017yef}. The relation between these scalar theories and the gauge and gravity theories can also be explained from another point of view: if we consider massive Yang-Mills and massive gravitational fields, the corresponding longitudinal modes are described by the non-linear sigma model and the (special\footnote{One can choose the parameters of massive gravity such that the resulting scalar field theory in the decoupling limit is the special Galileon.} \cite{Cachazo:2014xea,Cheung:2014dqa,Hinterbichler:2015pqa,Novotny:2016jkh} ) Galileon respectively \cite{Hinterbichler:2011tt,deRham:2014zqa}. This suggests the possibility of a broader relationship between these sets of theories. Indeed, it has been shown that there is a web of relationships between their scattering amplitudes \cite{Cachazo:2014xea,Cheung:2016prv,Cheung:2017ems,Cheung:2017yef,Zhou:2018wvn,Bollmann:2018edb}, see Fig.\ref{fig:web}. 

In this paper, we begin by analyzing the existence of a classical perturbative double copy for the NLSM radiation. The setup consists of point-particles weakly coupled to pions which evolve consistently with the NLSM field and whose deviations from their initial trajectories and color degrees of freedom are small. We assume that the NLSM coupling to the point-particles is invariant under the unbroken symmetry. This coupling is motivated by the fact that the NLSM can arise as the longitudinal mode of a massive Yang-Mills field which gives rise to pion couplings invariant under the unbroken symmetry. Similarly, we assume that the special Galileon couples through a conformal transformation which is motivated by the coupling that would arise in the decoupling limit of massive gravity for the Galileons. In addition to the double copy relation between these theories, it is also expected that one can perform a color-kinematics replacement from the bi-adjoint scalar and obtain the NLSM radiation as in the Yang-Mills case. Given this, we will also consider the zeroth copy case where the point-particles couple to the bi-adjoint scalar field, thus spanning the entire RHS of Fig. \ref{fig:web}. 

The observable that we want to map between theories is the radiation amplitude at spatial infinity $|\vec{x}|\rightarrow\infty$. For example, the on-shell radiation amplitude for the bi-adjoint scalar $\phi^{a\,\tilde{a}}$ is defined as 
\begin{equation}
\mathcal{A}^{a\,\tilde{a}}(k)= y \,  \mathcal{J}^{a\,\tilde{a}}(k)\Big|_{k^2=0},
\end{equation}
where the on-shell current $\mathcal{J}^{a\,\tilde{a}}(k)\Big|_{k^2=0}$ gives the flux of energy-momentum, color and angular momentum at spatial infinity, and is defined by the equations of motion $\square\phi^{a\,\tilde{a}} = y \, \mathcal{J}^{a\,\tilde{a}} $, with coupling constant $y$. Similar definitions hold for the non-linear sigma model and the special Galileon. In 4d, the probability of emission of a scalar can be written as \cite{Goldberger:2007hy}
\begin{equation}
\ud\text{Prob.}(0 \to \phi^{a\,\tilde{a}}) =|\mathcal{A}^{a\,\tilde{a}}|^2\frac{\ud \mathbf{k}}{(2\pi)^3}\frac{1}{2|\mathbf{k}|} \ .
\end{equation}
As the observation time grows, $T\rightarrow\infty$, the differential radiated power is given by
\begin{equation}
\frac{\ud P}{\ud \Omega \ud |\mathbf{k}|}=\left|\mathcal{A}^{a\,\tilde{a}}\right|^2 \frac{|\mathbf{k}|^2}{2(2\pi)^2 \, T}\ .
\end{equation}
The final goal is to be able to map the scalar radiation power emitted by a set of point-particles among different theories. In order to do so, we will only need to map between on-shell currents.

\subsection{Summary of results}

In this section, we summarize the procedure for obtaining the classical double copy for the radiation of scalar modes.  We show that by applying a special set of color-kinematics replacements, it is possible to transform the radiation field generated by point-particles interacting through a bi-adjoint scalar field to the one in which these particles interact through a non-linear sigma model field. Similarly, one can act on the NLSM radiation field to obtain the equivalent object for the double copy, {\it i.e..}, the special Galileon radiation. We consider the case where the impact parameters of the particles are large, and thus the particle number is conserved, since no particles are created or annihilated. The large separation of the particles accounts for the consistency of the perturbative calculation, a point that will be made more precise in the body of the paper. A crucial fact for the existence of the double copy is that the couplings of the scalar fields to the point-particles have the same coupling strength as the self-interactions of such fields. This is similar to the case of Yang-Mills and gravity.

For each theory, the point-particles carry different degrees of freedom depending on the couplings being considered. In the bi-adjoint scalar field case, the point-particles carry two color charges, $c^a$ and $\tilde{c}^{\tilde{a}}$ , each in the adjoint representation of the groups $G$ and $\tilde{G}$. In the case of the NLSM corresponding to the symmetry breaking pattern $G_L \times G_R \rightarrow G$ (with $G$ the diagonal subgroup), we will consider a coupling to the point-particles that is manifestly invariant under the unbroken symmetries. This means that the coupling will involve the ``covariant derivative'' of the Goldstone modes, $\nabla^\mu\phi^a \equiv f^{abc}(U^{-1}\partial_\mu U)_{bc}$, which in our case will couple to the color dipole moment ${M^a}^\mu$ of the point-particles. Manifest invariance under $G$ is sufficient to ensure invariance under the full group $G_L \times G_R$~\cite{Weinberg:1996kr}. Finally, the special Galileon coupling we will be using follows from a conformal transformation $g_{\mu\nu}\rightarrow(1+2\pi/\Lambda)g_{\mu\nu}$ of the point-particle action.  This transformation is motivated by the one implemented in massive gravity to remove the kinetic mixing between the helicity-2 modes and the longitudinal mode before taking the decoupling limit, in that case $\Lambda=M_{Pl}$.

Because we have different degrees of freedom carrying a color index in the bi-adjoint scalar and the non-linear sigma model, we will also need a replacement rule to map one to the other. Thus, for the single copy we need not only the usual color-kinematics replacements, which schematically are of the form 
\begin{equation}
\tilde{C}(\tilde{c}^{\tilde{a}})\rightarrow \tilde{N}(\{q\}) \ ,
\end{equation}
but also the color-color replacements
\begin{equation}
C(\{q\};c^a)\rightarrow C(\{q\}\cdot M^a) \ ,
\end{equation}
where $\{ q \}$ stands for the collection of momenta involved in the process. At second order in the couplings, the single copy color-color replacements are given by Eq.\eqref{CC2} and the color-kinematics by Eq.\eqref{CK2}. These replacements map the on-shell current Eq.\eqref{J2BS} into the on-shell current Eq.\eqref{JNLSM2der}.  At quartic order, the color-color replacements are given by Eq.\eqref{CC4} and the color-kinematics by Eq.\eqref{CK4}, these give a map between the on-shell currents in Eq.\eqref{J4BS} and Eq.\eqref{J4NLSMder}. 

For the double copy case instead we can simply perform a color-kinematics replacement of the form
\begin{equation}
C(\{q\}\cdot M^a)\rightarrow N(\{q\}) \ .
\end{equation}
The replacement rules at second order are found in Eq.\eqref{DCK2}, and the ones at quartic order in Eq.\eqref{DCK4}. These replacements create a map between on-shell currents: Eq.\eqref{JNLSM2der} maps onto Eq.\eqref{J2SG}, and Eq.\eqref{J4NLSMder} maps onto Eq.\eqref{J4SG}.  There are four main features that is worth highlighting about these replacements:
\begin{enumerate}
	\item {\bf Coupling constants:} the coupling constants in the three different theories are mapped into each other as follows:
	\begin{equation}
		y \to \frac{\sqrt{2}}{F} \to \frac{1}{\Lambda} \ .
	\end{equation}
	Thus, a result obtained in the biadjoint case with a precision of $\mathcal{O} (y^n)$ will be mapped onto an equivalent result for NLSM and special Galileon  at order $\mathcal{O} (1/F^n)$ and $\mathcal{O} (1/\Lambda^n)$ respectively. For the sake of brevity, from now on we will denote this level of precision as $\mathcal{O}(n)$.
	\item {\bf Color charges:} The color charges and dipole moments are mapped as
	\begin{equation*}
	\tilde{c}^{\tilde{a}} c^a\rightarrow q\cdot M^a \rightarrow 1 \ ,
	\end{equation*}
	where $q$ represents different momentum factors, depending on the specific color structure. This can be compared to the Yang-Mills-gravity case where the replacement is $c^a\rightarrow p^\mu$. In this case, we are mapping between scalar theories so no new structure with a Lorentz index appears uncontracted.
	\item {\bf Three-point vertex:} Color factors which involve only one structure constant are mapped to zero,
	\begin{equation*}
	f\cdot c \cdot c \rightarrow 0 .
	\end{equation*}
	In the gravitational double copy the color factor of the Yang-Mills three-point function, $f^{abc}$, is mapped to the color-stripped Yang-Mills three-point vertex. This is motivated by the BCJ double copy where one replaces the Yang-Mills color factor by a second copy of the Yang-Mills kinematic factor in order to obtain a gravitational amplitude. In the present case, the NLSM does not have a cubic vertex and thus the above color structure is mapped to zero.
	\item {\bf Color-kinematics duality for the double copy:}  The replacement rules that take the NLSM four-point amplitude color factor to the NLSM four point amplitude, {\it i.e..}
	\begin{equation*}
	i \,4\sqrt{2}\,  f^{abc}f^{bde} (q_\beta\cdot M_\beta)^d (q_\gamma\cdot M_\gamma)^e(q_\alpha\cdot M_\alpha)^c \rightarrow \frac{(q_\beta+q_\alpha)^2-(q_\gamma+q_\alpha)^2}{3} \ ,
	\end{equation*}
	maps color factors satisfying the Jacobi identity 
	\begin{equation*}
	\sum_\text{cyclic}f^{abc}f^{bde} (q_\beta\cdot M_\beta)^d (q_\gamma\cdot M_\gamma)^e(q_\alpha\cdot M_\alpha)^c = 0 \ ,
	\end{equation*}
	to kinematic factors that satisfy another Jacobi identity
	\begin{equation*}
	\sum_\text{cyclic}((q_\beta+q_\alpha)^2-(q_\gamma+q_\alpha)^2)=0 \ .
	\end{equation*}
	This provides a new example of the color-kinematics duality at the classical level. The analogue case for the gravitational double copy was studied in \cite{Shen:2018ebu}.
\end{enumerate}

\begin{figure}[!t]
	\includegraphics[scale=1]{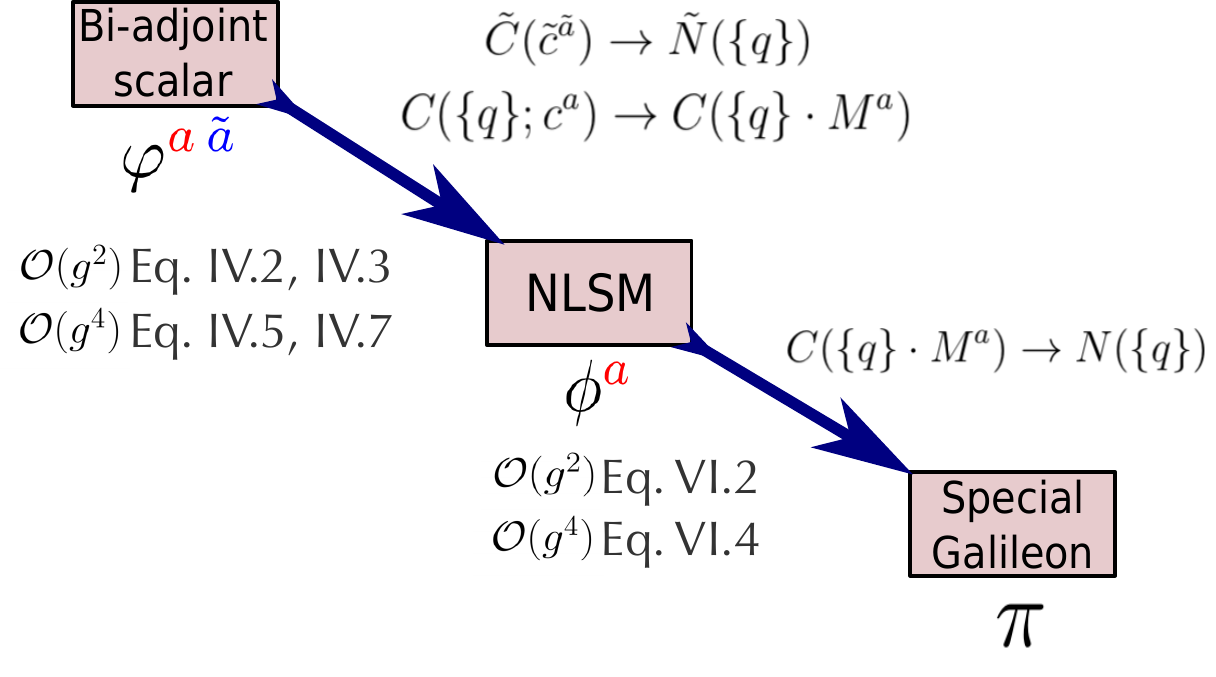}
	\caption{Summary of color-kinematics replacements used in this paper.}
	\label{fig:Rep}
\end{figure}

In the following, we carry out the perturbative calculation for each theory in detail. In section \ref{BS}, we compute the bi-adjoint scalar radiation and in section \ref{NLSM} that for the non-linear sigma model case. In section \ref{singlecopy} we then explain the color-kinematics and color-color replacements that transform the bi-adjoint scalar result into the NLSM one. We continue in section \ref{SG} with the calculation of the special Galileon radiation and in section \ref{doublecopy} derive the color-kinematics replacements that lead to the double copy. We conclude in section \ref{conclusions}.

\section{Bi-adjoint scalar radiation} \label{BS}
In this section, we compute the radiation field produced by color charges coupled through the bi-adjoint scalar field. The $\mathcal{O}(2)$ result was first computed in \cite{Goldberger:2017frp} and extended to order $\mathcal{O}(4)$ in \cite{Shen:2018ebu}. In the following, we show these results for completeness while clarifying some technical details of the calculation.

The bi-adjoint scalar field transforms in the adjoint representation of the group $G\times\tilde{G}$ and has cubic interactions, described by the Lagrangian
\begin{equation}
\lag_{BS}=\frac{1}{2}\left(\partial\varphi^{a\,\tilde{a}}\right)^2-\frac{y}{3}\,f^{abc}\,\tilde{f}^{\tilde{a}\tilde{b}\tilde{c}}\,\varphi^{a\,\tilde{a}}\varphi^{b\,\tilde{b}}\varphi^{c\,\tilde{c}} \ .
\end{equation}
Our goal is to compute perturbatively the scalar radiation field generated by a set of color charges coming from infinity, which will evolve consistently together with the field they generate. The point-particles carry color charges also transforming in the adjoint representation of $G\times\tilde{G}$ and move along the worldlines $x^{\mu}_{\alpha}(\lambda)$, where $\lambda$ is the coordinate along the worldline, and $\alpha$ labels the individual particles. These point-particles are coupled to the scalar field in the following way: 
\begin{equation}
S_\text{pp}=-\frac{1}{2}\sum_{\alpha}\int \ud \lambda \left[\eta^{-1}(\lambda)\,\frac{\ud x_\alpha}{\ud \lambda}\cdot\frac{\ud x_\alpha}{\ud \lambda}+\eta(\lambda)\left(m_\alpha^2-2\,y\, \varphi^{a \, \tilde{a}}(x_\alpha) \, c^a_\alpha(\lambda)\tilde{c}^{\tilde{a}}_\alpha(\lambda)\right)\right] \ ,
\end{equation}
where the einbein $\eta(\lambda)$ is a Lagrange multiplier that ensures invariance under reparametrizations of $\lambda$, and $c^a$ and $\tilde{c}^{\tilde{a}}$ are color charges transforming in the adjoint representations of $G$ and $\tilde{G}$ respectively. For the purpose of this paper, the specific Lagrangian realization giving rise to the color charges is not relevant and thus is not considered here, but a discussion regarding this can be found in \cite{Goldberger:2017frp}.  The total color currents are
\begin{equation}
J^{\mu, \, a}=J^{\mu, \, a}_\text{N. BS}+J^{\mu, \, a}_\text{pp} \ , \quad \quad J^{\mu, \, \tilde{a}}=J^{\mu, \, \tilde{a}}_\text{N. BS}+J^{\mu, \, \tilde{a}}_\text{pp} \ .
\end{equation}
Here, $J^{\mu, \, a}_\text{N. BS}$ and $J^{\mu, \, \tilde{a}}_\text{N. BS}$ are the Noether currents derived from $\lag_\text{BS}$ due to the invariance under $G$ and $\tilde{G}$ and read
\begin{equation}
J^{\mu, \, a}_\text{N. BS}= f^{abc} \varphi^{b\, \tilde{b}} \, \partial^\mu \varphi^{c \, \tilde{b}} \ , \quad\quad J^{\mu, \, \tilde{a}}_\text{N. BS}= \tilde{f}^{\tilde{a}\tilde{b}\tilde{c}} \varphi^{b\, \tilde{b}} \, \partial^\mu \varphi^{b \, \tilde{c}}\ , 
\end{equation}
while the leading order currents produced by the point-particles are given by 
\begin{equation}
J^{\mu, \, a}_\text{pp}=\sum_{\alpha}\int\ud\lambda \, c_\alpha^a(\lambda)v^\mu_\alpha \, \delta^d(x-x^\alpha(\lambda)) \ , \quad \,
J^{\mu, \, \tilde{a}}_\text{pp}=\sum_{\alpha}\int\ud\lambda \, \tilde{c}_\alpha^{\tilde{a}}(\lambda)v^\mu_\alpha \, \delta^d(x-x^\alpha(\lambda)) \ .
\end{equation}
where $v^\mu_\alpha$ is the velocity of the point-particle $\alpha$ carrying color charge $c_\alpha^a(\lambda)$ or $\tilde{c}_\alpha^{\tilde{a}}(\lambda)$. The next to leading order contributions to these currents include finite size effects. By varying the action $S_\text{BS}+S_\text{pp}$ and considering current conservation, we obtain the equations of motion for the coordinates and the color charges
\begin{align}
&\frac{\ud p^\mu_\alpha}{\ud s}+ y \, c^a_\alpha(s) \tilde{c}^{\tilde{a}}_\alpha(s) \partial^\mu\varphi_{a \, \tilde{a}}(x^\mu_\alpha)=0 \ , \label{ppBS} \\
\frac{\ud c^a_\alpha}{\ud s}+ y \, f^{abc} c^c_\alpha(s) & \tilde{c}^{\tilde{b}}_\alpha(s) \varphi^{b \, \tilde{b}}(x^\mu_\alpha)=0 \ , \quad  \frac{\ud \tilde{c}^{\tilde{a}}_\alpha}{\ud s}+ y \, \tilde{f}^{\tilde{a}\tilde{b}\tilde{c}}  \tilde{c}^{\tilde{c}}_\alpha(s) c^b_\alpha(s)\varphi^{b \, \tilde{b}}(x^\mu_\alpha)=0 \ , \label{ccBS}
\end{align}
where $p^\mu_\alpha\equiv\ud x_\alpha^\mu / \ud s$ is the momentum of the particle $\alpha$ and $\ud s=\eta \, \ud \lambda$. 
 
\subsection{Perturbative solutions}
The equation motion for the bi-adjoint scalar field can be written as 
\begin{equation}
\square\varphi^{a\,\tilde{a}} = y \, \mathcal{J}^{a\,\tilde{a}} \ ,
\end{equation}
where the source current is
\begin{equation}
\mathcal{J}^{a\,\tilde{a}} =-f^{abc}\,\tilde{f}^{\tilde{a}\tilde{b}\tilde{c}}\,\varphi^{b\,\tilde{b}}\varphi^{c\,\tilde{c}} +\sum_{\alpha}\int\ud s c^a_\alpha(\lambda)\tilde{c}^{\tilde{a}}_\alpha(\lambda) \delta^d(x-x_\alpha(s)) \ .
\end{equation}
This allows us to compute the radiation field at $|\vec{x}|\rightarrow\infty$ in terms of the Fourier transform of the source:
\begin{equation}
\varphi^{a\,\tilde{a}}=y\int\frac{\ud^d k}{(2\pi)^d}\, \frac{e^{-i\,k\cdot x}}{k^2} \mathcal{J}^{a\,\tilde{a}}(k) \ .
\end{equation}
The initial configuration consists of $N$ charged particles that are moving with constant velocity at $s=-\infty$. Thus, the initial conditions for the color-charged point-particles are:
\begin{align}
x^\mu&_\alpha|_{s\rightarrow-\infty}=b^\mu_\alpha+p^\mu_\alpha s \ ,\\
c^a_\alpha|_{s\rightarrow-\infty}&=c^a_\alpha \ ,  \quad \quad \tilde{c}^{\tilde{a}}_\alpha|_{s\rightarrow-\infty}=\tilde{c}^{\tilde{a}}_\alpha \ ,
\end{align}
where $b^\mu_\alpha$ are the (spacelike) impact parameters.  

In what follows, we compute the solutions perturbatively in powers of the coupling strength. The actual dimensionless parameter that controls the expansion is a combination of the coupling strength and kinematic factors, given by~\cite{Goldberger:2016iau}
\begin{equation*}
\epsilon\propto y^2\frac{c^2\, \tilde{c}^2}{E^3\,b} \ ,
\end{equation*}
where $E\gtrsim m$ is the energy of the point-particle and $b$ is its impact parameter. In this expression we have neglected the phase space volume. Notice that the perturbation parameter is inversely proportional to the impact parameter. This is consistent with our set up of particles that are far apart from each other and which only experience small deviations as they interact through the scalar field. Indeed, the fact that $\epsilon\ll 1$ ensures that the deviations are small compared to the impact parameter.  We can now find the $\mathcal{O}(1)$ field 
\begin{align}
\varphi^{a\,\tilde{a}}\big{|}_{\mathcal{O}(1)}&=-y\int\frac{\ud^d k}{(2\pi)^d}\, \frac{e^{-i\,k\cdot x}}{k^2}  \mathcal{J}^{a\,\tilde{a}}(k) \big{|}_{\mathcal{O}(0)} \nonumber \\ 
&=-y\sum_{\alpha}\int\frac{\ud^d k}{(2\pi)^d}\, c_\alpha^a\tilde{c}^{\tilde{a}}_\alpha \frac{e^{-i\,k\cdot\left(x-b_\alpha\right)}}{k^2} \, 2\pi \delta(k\cdot p_\alpha) \ . 
\end{align}
Notice that on-shell ($k^2=0$) the field vanishes unless $k^\mu\propto p^\mu$ but $p^\mu$ is timelike, therefore there is no radiation at this order, as we should expect, since static point-particles do not radiate.
\begin{figure}[!t]
	\includegraphics[scale=1]{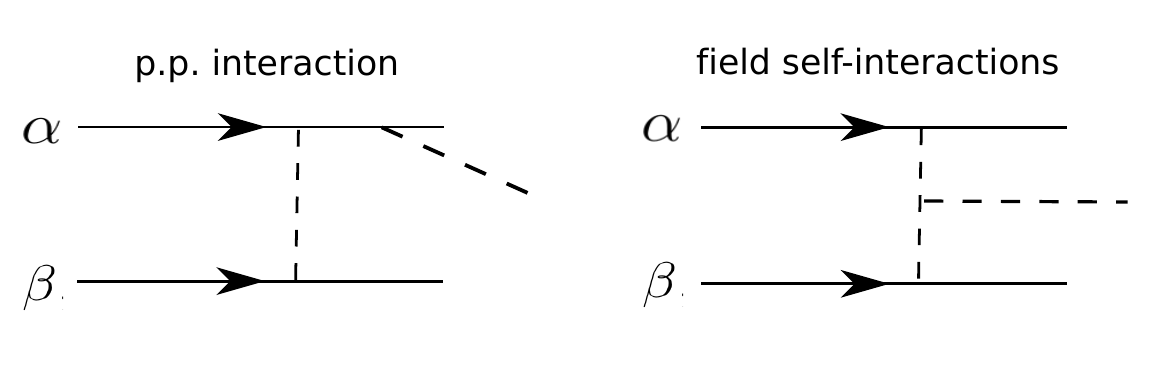}\caption{Interactions contributing to the current at $\mathcal{O}(2)$. The scalar field theory for the right-hand side graph has cubic self-interactions. If this were not the case, such graph would not contribute to $\mathcal{J}|_{\mathcal{O}(2)}$. The solid lines are the worldlines of the point-particles (which can carry color charge degrees of freedom depending on the theory under consideration) and the dashed lines represent the scalar field.}
	\label{order2}
\end{figure}
We now proceed to obtain the next order perturbation for the deviations of the point-particle trajectories and color charges. These are obtained by considering 
\begin{align}
&x^\mu_\alpha=b^\mu_\alpha+p^\mu_\alpha s+ \bar{x}_\alpha^\mu(s) \ ,\\
c^a_\alpha=c^a_\alpha&+\bar{c}^a_\alpha(s) \ , \quad \quad \tilde{c}^{\tilde{a}}_\alpha=\tilde{c}^{\tilde{a}}_\alpha+\bar{\tilde{c}}^{\tilde{a}}_\alpha(s)\ ,
\end{align}
where the barred quantities vanish at $s=-\infty$. Substituting the $\mathcal{O}(1)$ field into the equations of motion \eqref{ppBS} and \eqref{ccBS} we find
\begin{align}
\bar x^\mu_\alpha\big{|}_{\mathcal{O}(2)}&= i  y^2\,\sum_{\beta\neq\alpha} c_\alpha\cdot c_\beta \,\, \tilde{c}_\alpha\cdot \tilde{c}_\beta \, \int\frac{\ud^d q}{(2\pi)^d}\, q^\mu\frac{e^{-i\,q\cdot\left(b_{\alpha\beta}+p_\alpha s\right)}}{q^2 \, (q\cdot p_\alpha)^2} \, 2\pi\, \delta(q\cdot p_\beta) \ , \label{x2BS} \\
\bar c_\alpha^a\big{|}_{\mathcal{O}(2)}&=i y^2 f^{abc} \sum_{\beta\neq\alpha} c_\alpha^c c_\beta^b \, \, \tilde{c}_\alpha\cdot \tilde{c}_\beta\int\frac{\ud^d q}{(2\pi)^d}\,  \frac{e^{-i\,q\cdot\left(b_{\alpha\beta}+p_\alpha s\right)}}{q^2(q\cdot p_\alpha)} \, 2\pi\, \delta(q\cdot p_\beta)  \ , \label{c2BS} \\
\tilde{c}_\alpha^{\tilde{a}}\big{|}_{\mathcal{O}(2)}&=i y^2 \tilde{f}^{\tilde{a}\tilde{b}\tilde{c}} \sum_{\beta\neq\alpha} \tilde{c}_\alpha^{\tilde{c}} \tilde{c}_\beta^{\tilde{b}} \, \, c_\alpha\cdot c_\beta\int\frac{\ud^d q}{(2\pi)^d}\,  \frac{e^{-i\,q\cdot\left(b_{\alpha\beta}+p_\alpha s\right)}}{q^2(q\cdot p_\alpha)} \, 2\pi\, \delta(q\cdot p_\beta)  \ , \label{c2tildeBS}
\end{align}
where $b_{\alpha\beta}\equiv b_\alpha-b_\beta$. The source $\mathcal{J}^{a\,\tilde{a}}(k)$ at $\mathcal{O}(y^{2})$ is given by 
\begin{align}
\mathcal{J}^{a\,\tilde{a}}(k) \big{|}_{\mathcal{O}(2)}=&\sum_{\alpha}\int \ud s\, e^{i\,k\cdot\left(b_\alpha+p_\alpha s \right)}\Big[ik\cdot \bar{x}_\alpha(s)\big{|}_{\mathcal{O}(2)}c^a_\alpha\tilde{c}_\alpha^{\tilde{a}}+\bar{c}^a_\alpha(s)\big{|}_{\mathcal{O}(2)}\tilde{c}_\alpha^{\tilde{a}}\nonumber \\
&+c^a_\alpha(s)\bar{\tilde{c}}_\alpha^{\tilde{a}}\big{|}_{\mathcal{O}(2)}\Big] -f^{abc}\,\tilde{f}^{\tilde{a}\tilde{b}\tilde{c}}\,\int\ud^dx e^{ik\cdot x}\varphi^{b\,\tilde{b}}\big{|}_{\mathcal{O}(1)}\varphi^{c\,\tilde{c}}\big{|}_{\mathcal{O}(1)} 
\ . \label{J2BSa}
\end{align}
After using our previous results, the source current becomes
\begin{equation}
\mathcal{J}^{a\,\tilde{a}}(k) \big{|}_{\mathcal{O}(2)}=y^2\sum_{\alpha, \, \beta\neq\alpha}\int_{q_\alpha, \, q_\beta}\, \rho^{a\,\tilde{a}}_{\alpha\beta}(k) \mu_{\alpha,\beta}(k) \ , \label{J2BS}
\end{equation}
\vspace{-0.3cm}
where
\begin{align}
\rho^{a\,\tilde{a}}_{\alpha\beta}(k)\equiv &-\frac{q_\alpha^2}{(k\cdot p_\alpha)}\Big[ c_\alpha^a (c_\alpha\cdot\!c_\beta) \, \, \tilde{c}_\alpha^{\tilde{a}} (\tilde{c}_\alpha\cdot\!\tilde{c}_\beta)\,  \frac{k\cdot q_\beta}{k\cdot p_\alpha}-i \, f^{abc} c_\alpha^c c_\beta^b \, \tilde{c}_\alpha^{\tilde{a}} (\tilde{c}_\alpha \cdot \tilde{c}_\beta)\, \nonumber \\
&-i\tilde{f}^{\tilde{a}\tilde{b}\tilde{c}}\tilde{c}_\alpha^{\tilde{c}}\tilde{c}_\beta^{\tilde{b}} c_\alpha^a (c_\alpha \cdot c_\beta) \, \Big]-f^{abc}\,\tilde{f}^{\tilde{a}\tilde{b}\tilde{c}}c_\alpha^bc_\beta^c\tilde{c}_\alpha^{\tilde{b}}\tilde{c}_\beta^{\tilde{c}} \ ,
\end{align}
with $\int_q\equiv\int\frac{\ud^d q}{(2\pi)^d}$, and
\begin{equation}
\mu_{\alpha,\beta}(k)\equiv (2\pi)\delta(q_\alpha\cdot p_\alpha)\frac{e^{i\,q_\alpha\cdot b_{\alpha}}}{q_\alpha^2}(2\pi) \delta(q_\beta\cdot p_\beta) \frac{e^{i\,q_\beta\cdot b_{\beta}}}{q_\beta^2}(2\pi)^d\delta^d(k-q_\beta-q_\alpha) \ . 
\end{equation}
The above result for the source current heavily relies on the use of the delta functions in $\mu_{\alpha,\beta}(k)$. This will be the case for all the final results that we present. One can think of this perturbative solution in terms of Feynman diagrams. At second order in the coupling, the contributions to the bi-adjoint current are given by the graphs in Fig. \ref{order2}. The first term in the parentheses in Eq.\eqref{J2BSa} corresponds to the graph on the left-hand side of Fig. \ref{order2}. This graph only shows the case of the scalar field radiated by particle $\alpha$, but we should also include the case where it is radiated from particle $\beta$. This is taken into account by the sum over point-particles. The last term of Eq.\eqref{J2BSa} comes from the graph on the right-hand side which corresponds to the self-interactions of the field.

\begin{figure}[!t]
	\includegraphics[scale=0.75]{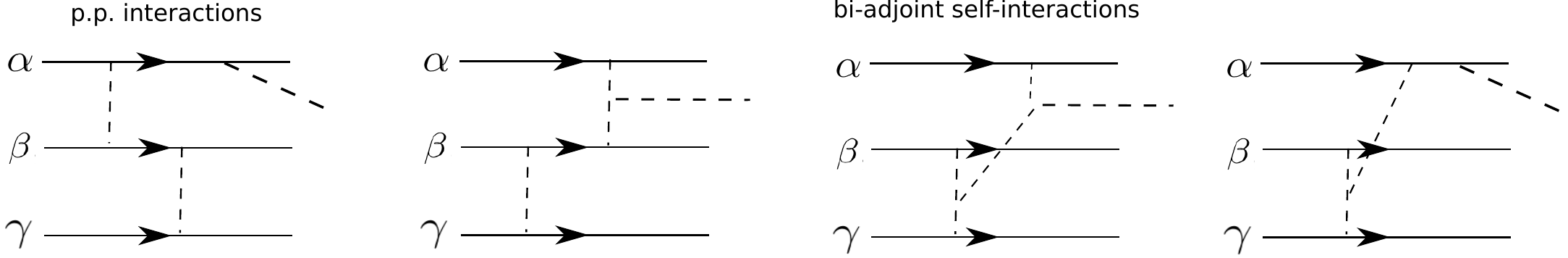}\caption{Interactions contributing to the current at $\mathcal{O}(4)$. The left-hand side shows the contribution from the interactions between the point-particles and the bi-adjoint scalar field. The center and right-hand side graphs show the contributions from the bi-adjoint scalar 3-point vertex. }
	\label{BSorder4}
\end{figure}

As we will see in the next section, the NLSM self-interactions will only contribute at next to leading order in perturbations. Hence, in order to construct a satisfactory copy we will compute the source current for the radiation field for the bi-adjoint scalar at $\mathcal{O}(4)$. The source at this order is given by 
\begin{align}
\mathcal{J}^{a\,\tilde{a}}(k)&\big{|}_{\mathcal{O}(4)}=-f^{abc}\,\tilde{f}^{\tilde{a}\tilde{b}\tilde{c}}\,\int\ud^dx e^{ik\cdot x}\varphi^{b\,\tilde{b}}\big{|}_{\mathcal{O}(1)}\varphi^{c\,\tilde{c}}\big{|}_{\mathcal{O}(3)} +\sum_{\alpha}\int \ud s\, e^{i\,k\cdot\left(b_\alpha+p_\alpha s \right)}\Bigg\{\bar{c}^a_\alpha(s)\big{|}_{\mathcal{O}(4)}\tilde{c}_\alpha^{\tilde{a}}\nonumber \\
& +c^a_\alpha(s)\bar{\tilde{c}}_\alpha^{\tilde{a}}\big{|}_{\mathcal{O}(4)}+\bar{c}^a_\alpha(s)\big{|}_{\mathcal{O}(2)}\bar{\tilde{c}}_\alpha^{\tilde{a}}\big{|}_{\mathcal{O}(2)}+i k\cdot \bar{x}_\alpha(s)\big{|}_{\mathcal{O}(2)}\left[\bar{c}^a_\alpha(s)\big{|}_{\mathcal{O}(2)}\tilde{c}_\alpha^{\tilde{a}}+c^a_\alpha(s)\bar{\tilde{c}}_\alpha^{\tilde{a}}\big{|}_{\mathcal{O}(2)}\right] \nonumber \\
&+i k\cdot \bar{x}_\alpha(s)\big{|}_{\mathcal{O}(4)}c^a_\alpha \tilde{c}_\alpha^{\tilde{a}} +\frac{1}{2}[ik\cdot \bar{x}_\alpha(s)\big{|}_{\mathcal{O}(2)}]^2c^a_\alpha\tilde{c}_\alpha^{\tilde{a}}\Bigg\}\ , \label{J4BS}
\end{align}
which corresponds to the graphs in Fig.\ref{BSorder4}. 
The term in curly brackets contains the deflections of the point-particle coordinates and color charges at next to leading order, which are given by 
\begin{align}
&\bar x^\mu_\alpha\big{|}_{\mathcal{O}(4)}=y^4 \underset{\gamma \neq\{\alpha,\beta\}}{\sum_{\beta \neq\alpha,}} \underset{q_\beta, \, q_\gamma\, q_\delta}{\int} \frac{ e^{-i\,q_\delta\cdot\left(b_{\alpha}+p_\alpha s\right)} }{\left( q_\delta\cdot p_\alpha\right)^2}\mu_{\beta,\gamma}(q_\delta) \Biggl\{ \frac{q_\gamma^\mu}{q_\beta\cdot p_\alpha}\Big[f^{abc}c_\alpha^bc_\beta^cc_\gamma^a (\tilde{c}_\alpha\cdot \tilde{c}_\beta)(\tilde{c}_\alpha\cdot \tilde{c}_\gamma) \nonumber\\
&+\tilde{f}^{\tilde{a}\tilde{b}\tilde{c}} \tilde{c}_\alpha^{\tilde{b}}\tilde{c}_\beta^{\tilde{c}}\tilde{c}_\gamma^{\tilde{a}}(c_\alpha\cdot c_\beta)(c_\alpha\cdot c_\gamma)+i\frac{q_\beta\cdot q_\gamma}{q_\beta\cdot p_\alpha}(c_\alpha\cdot c_\beta)(c_\alpha\cdot c_\gamma)(\tilde{c}_\alpha\cdot \tilde{c}_\beta)(\tilde{c}_\alpha\cdot \tilde{c}_\gamma) \Big]+\frac{iq_\delta^\mu}{q_\delta^2} \rho^{a\,\tilde{a}}_{\beta\gamma}(q_\delta)c_\alpha^a\tilde{c}_\alpha^{\tilde{a}}\Biggl\} \label{x4BS} \\
&\bar{c}_\alpha^a\big{|}_{\mathcal{O}(4)}=y^4f^{abc}\underset{\gamma \neq\{\alpha,\beta\}}{\sum_{\beta \neq\alpha,}} \underset{q_\beta, \, q_\gamma\, q_\delta}{\int} \frac{ e^{-i\,q_\delta\cdot\left(b_{\alpha}+p_\alpha s\right)} }{\left( q_\delta\cdot p_\alpha\right)}\mu_{\beta,\gamma}(q_\delta)\Biggl\{\frac{1}{q_\beta\cdot p_\alpha}\Big[f^{bde}c_\alpha^ec_\beta^dc_\gamma^c (\tilde{c}_\alpha\cdot \tilde{c}_\beta)(\tilde{c}_\alpha\cdot \tilde{c}_\gamma) \nonumber\\
&+\tilde{f}^{\tilde{b}\tilde{d}\tilde{e}} \tilde{c}_\alpha^{\tilde{e}}\tilde{c}_\beta^{\tilde{d}}\tilde{c}_\gamma^{\tilde{b}}(c_\alpha\cdot c_\beta)c_\alpha^b c_\gamma^c-i\frac{q_\beta\cdot q_\gamma}{q_\beta\cdot p_\alpha}c_\alpha^b c_\gamma^c(c_\alpha\cdot c_\beta)(\tilde{c}_\alpha\cdot \tilde{c}_\beta)(\tilde{c}_\alpha\cdot \tilde{c}_\gamma)\Big]+\frac{i}{q_\delta^2} \rho^{b\,\tilde{b}}_{\beta\gamma}(q_\delta)c_\alpha^c\tilde{c}_\alpha^{\tilde{b}}\Biggl\} \ , \label{c4BS}\\
&\bar{\tilde{c}}_\alpha^{\tilde{a}}\big{|}_{\mathcal{O}(4)}=y^4\tilde{f}^{\tilde{a}\tilde{b}\tilde{c}}\underset{\gamma \neq\{\alpha,\beta\}}{\sum_{\beta \neq\alpha,}} \underset{q_\beta, \, q_\gamma\, q_\delta}{\int} \frac{ e^{-i\,q_\delta\cdot\left(b_{\alpha}+p_\alpha s\right)} }{\left( q_\delta\cdot p_\alpha\right)}\mu_{\beta,\gamma}(q_\delta)\Biggl\{\frac{1}{q_\beta\cdot p_\alpha}\Big[f^{bde}c_\alpha^ec_\beta^dc_\gamma^b(\tilde{c}_\alpha\cdot \tilde{c}_\beta)\tilde{b}_\alpha^{\tilde{c}}\ \tilde{c}_\gamma^{\tilde{c}} \nonumber\\
&+\tilde{f}^{\tilde{b}\tilde{d}\tilde{e}} \tilde{c}_\alpha^{\tilde{e}}\tilde{c}_\beta^{\tilde{d}}\tilde{c}_\gamma^{\tilde{c}}(c_\alpha\cdot c_\beta)(c_\alpha \cdot  c_\gamma)-i\frac{q_\beta\cdot q_\gamma}{q_\beta\cdot p_\alpha}(c_\alpha\cdot c_\beta)(c_\alpha\cdot c_\gamma)(\tilde{c}_\alpha\cdot \tilde{c}_\beta)\tilde{c}_\alpha^{\tilde{b}} \tilde{c}_\gamma^{\tilde{c}}\Big]+\frac{i}{q_\delta^2} \rho^{b\,\tilde{b}}_{\beta\gamma}(q_\delta)\tilde{c}_\alpha^{\tilde{c}}c_\alpha^{b}\Biggl\} \ , \label{c4tildeBS}
\end{align}
Notice that the momentum involved in the propagators that appear in these calculations corresponds to the momentum exchanged with the point-particle. After using the fact that
\begin{equation}
\varphi^{a\,\tilde{a}}\big{|}_{\mathcal{O}(3)}=-y\int\frac{\ud^d k}{(2\pi)^d}\, \frac{e^{-i\,k\cdot x}}{k^2}  \mathcal{J}^{a\,\tilde{a}}(k) \big{|}_{\mathcal{O}(2)} \ ,
\end{equation}
with $\mathcal{J}^{a\,\tilde{a}}(k) \big{|}_{\mathcal{O}(2)}$ given by Eq.\eqref{J2BS}, the first term, which comes from the field self-interactions, reads
\begin{equation} \label{J 4 biadjoint}
\mathcal{J}^{a\,\tilde{a}}(k)\big{|}_{\mathcal{O}(4)}\supset -2y^4\underset{\gamma \neq\{\alpha,\beta\}}{\sum_{\alpha,\,\beta \neq\alpha,}} \underset{q_\alpha,\,q_\beta, \, q_\gamma\, q_\delta}{\int} f^{abc}\,\tilde{f}^{\tilde{a}\tilde{b}\tilde{c}} c_\alpha^b \tilde{c}_\alpha^{\tilde{b}} \,\frac{\rho^{c\,\tilde{c}}_{\beta\gamma}(q_\delta)}{q_\delta^2} \,  \mu_{\alpha,\beta,\gamma}(k) \,(2\pi)^d\delta^d(q_\delta-q_\beta-q_\gamma) \ ,
\end{equation}
where $\mu_{\alpha,\beta,\gamma}(k)$ is the straightforward  generalization of $\mu_{\alpha ,\beta}(k)$, namely
\begin{equation}
\mu_{\alpha,\beta, \gamma}(k)\equiv (2\pi)\delta(q_\alpha\cdot p_\alpha)\frac{e^{i\,q_\alpha\cdot b_{\alpha}}}{q_\alpha^2}(2\pi) \delta(q_\beta\cdot p_\beta) \frac{e^{i\,q_\beta\cdot b_{\beta}}}{q_\beta^2} (2\pi)\delta(q_\gamma\cdot p_\gamma)\frac{e^{i\,q_\gamma\cdot b_{\gamma}}}{q_\gamma^2} (2\pi)^d\delta^d(k-q_\beta-q_\alpha-q_\gamma)\ .
\end{equation}
Notice that the current $\mathcal{J}^{a\,\tilde{a}}(k)$ should be symmetric under interchange of particles. This symmetry is not manifest in Eq. \eqref{J 4 biadjoint}, but it is realized by the sum over the particle indices $\alpha, \beta$ and $\gamma$.

\section{Non-linear sigma model radiation}\label{NLSM}

Consider now the non-linear sigma model (NLSM) based on the simple compact Lie group $G$; that is, the model corresponding to the symmetry breaking $G_L \times G_R \rightarrow G_\text{diag}$, where $G_L= G_R= G_\text{diag}\equiv G$. The leading order effective Lagrangian is given by
\begin{equation}
\lag^{(2)}_\text{NLSM}=\frac{F^{2}}{4}\text{Tr}\left( \partial_\mu U \partial^\mu U^{-1} \right) \ ,
\end{equation}
where $U=g_Rg_L^{-1}$, and $g_{R(L)}$ is an element of the group $G_{R(L)}$. We will use the exponential parametrization
\begin{equation}
U=e^{i\frac{\sqrt{2}}{F}\phi^a T^a} \ ,
\end{equation}
where $\phi^a$ are the Goldstone boson fields and $T^a$ are the generators of $G$. Given that the pattern $G_L \times G_R \rightarrow G_\text{diag}$ is a simple generalization of the one describing QCD pions, in what follows we will often refer to the NLSM fields simply as pions. Since all quantities with a color index will transform in the adjoint representation, we'll find it convenient to follow the conventions that are often adopted in the amplitudes literature (see {\it e.g.}~\cite{Kampf:2012fn,Kampf:2013vha,Chen:2013fya,Du:2016tbc}). Hence, our generators satisfy the following relations:
\begin{equation}
\text{Tr}\left(T^a T^b\right)=\delta^{a b}, \qquad\quad  [T^a,T^b]=i\sqrt{2}f^{abc}T^c, \qquad\quad  (T^a)^{bc}=-i\sqrt{2}f^{abc} \ .
\end{equation}
With this parametrization, the strength of self-interactions is determined by the coupling $\sqrt{2}/F$. In terms of the Goldstone  fields, the Lagrangian can be rewritten as
\begin{equation}
\lag^{(2)}=-\partial\phi^T\cdot G(\phi)\cdot\partial\phi \ ,
\end{equation}
where we have defined
\begin{align}
G(\phi)\equiv \sum_{n=1}^{\infty}\frac{(-1)^n}{(2n)!}\left(\frac{2 D_\phi}{F}\right)^{2n-2} \ , \qquad \qquad 
D_\phi^{ab}\equiv-if^{abc}\phi^c \ .
\end{align} 
In this case, we want to consider a coupling to the point-particle that preserves the unbroken symmetry $G$, which means that it involves the pion covariant derivative $\nabla_\mu\phi^a\equiv f^{abc}(U^{-1}\partial_\mu U)_{bc}$. Consider a coupling to a dipole moment $M^a_\mu(\lambda)$ localized on the worldline:
\begin{equation}
S_\text{pp}=-\frac{1}{2}\sum_{\alpha}\int \ud \lambda \left[\eta^{-1}(\lambda)\,\frac{\ud x_\alpha}{\ud \lambda}\cdot\frac{\ud x_\alpha}{\ud \lambda}+\eta(\lambda)m_\alpha^2\left(1-2 {M^a_\alpha}^\mu(\lambda) \,\nabla_\mu \phi^a\right)\right] \ , \label{SppNLSM2}
\end{equation}
where the pion covariant derivative in the exponential parametrization is
\begin{equation}
\nabla_\mu\phi^a=\frac{2}{F}\partial_\mu\phi^a-\frac{2}{F^2}f^{abc} \partial_\mu\phi^b \, \phi^c+\frac{4}{3 F^3}f^{abc}f^{bde}\phi^c\phi^e\partial_\mu\phi^d+\cdots \ .
\end{equation}
From this coupling, we can read off the current generated by the color charges. Up to next to next leading order this current is 
\begin{equation}
J^{\mu, \, a}_\text{pp}=\sqrt{2}\sum_{\alpha}m_\alpha^2\int\ud s \, \left({M^a_\alpha}^\mu(s)+\frac{1}{F}f^{abc}{M^b_\alpha}^\mu(s)\, \phi^c+\frac{2}{3F^2}f^{abc}f^{bde}\phi^c\phi^e{M^d_\alpha}^\mu\right)\, \delta^d(x-x^\alpha(\lambda)) \ .
\end{equation}
We obtain the equation of motion that determines the evolution of the point-particle coordinates $x^{\mu}_{\alpha}$ by varying the point-particle action $S_\text{pp}$. At next to leading order in the coupling this yields
\begin{equation}
\frac{\ud p^\mu_\alpha}{\ud s}+\frac{2}{F}m_\alpha^2{M^a_\alpha}^\nu\partial^\mu\left(\partial_\nu\phi^a-\frac{1}{F}f^{abc} \partial_\nu\phi^b \, \phi^c+\cdots\right)=0 \ . \label{ppDip}
\end{equation}
Similarly, we obtain the equations of motion determining the evolution of the dipole moment from the conservation of the total color current
\begin{equation}
J^{\mu, \, a}=J^{\mu, \, a}_\text{N. NLSM}+J^{\mu, \, a}_\text{pp} \ .
\end{equation}
Here, $J^{\mu, \, a}_\text{N. NLSM}$ is the Noether current derived from $\lag^{(2)}_\text{NLSM}$ and reads
\begin{equation}
J^{\mu, \, a}_\text{N. NLSM}= \sqrt{2}f^{abc} \phi^b \, G(\phi) \, \partial^\mu \phi^c \ .
\end{equation}
Given this, $\partial_\mu J^{\mu, \, a}=0$ implies that the dipole evolves according to
\begin{equation}
 k\cdot{M^a_\alpha}^\mu=-\frac{2i}{F}f^{abc}\partial_\mu\phi^c({M^b_\alpha}^\mu+\frac{1}{F}f^{bde}{M^d_\alpha}^\mu\phi^e)+\frac{1}{F}f^{abc}k_\mu\phi^c({M^b_\alpha}^\mu+\frac{4}{3F}f^{bde}{M^d_\alpha}^\mu\phi^e) \ . \label{ccDerCDip}
\end{equation}
The above equation is found after performing a  Fourier transformation and using the equation of motion obtained from varying $S_\text{NLSM}+S_\text{pp}$ that reads
\begin{align}
2\,G(\phi)\square\phi^a+2\partial G(\phi)\cdot\partial\phi^a-\partial\phi^T\cdot \partial_\phi G(\phi)^a\cdot\partial\phi-\frac{\sqrt{2}}{F}\partial_\mu J^{\mu, \, a}_\text{pp}=0 \ . \label{NLSMeomDip}
\end{align}

\subsection{Perturbative solutions}
We proceed to find an expression for the radiation field $\phi^a$ produced due to the interactions of the point-particles. We will again assume that the point-particles are well separated and that the impact parameters are large, so that we can construct a perturbative solution in powers of the NLSM coupling. As in the previous case,  the actual perturbation parameter is a combination of the coupling strength and kinematic factors, given by
\begin{equation*}
\epsilon\propto \frac{1}{F^{2}}\frac{E \, (k\cdot M_\alpha)\cdot(k\cdot M_\beta)}{b_{\alpha\beta}} \ ,
\end{equation*} 
where $k=\mathcal{O}(E)$ and the dipole is $M^a\propto c^a d$, where $d\ll\bar{x}$ is a measure of the size of the point-particle. We start by rewriting the NLSM equation of motion \eqref{NLSMeomDip} as
\begin{equation}
\square\phi^a=\frac{\sqrt{2}}{F}\mathcal{J}^a \ ,
\end{equation}
where the source current $\mathcal{J}^a$ is defined as
\begin{align}
&\mathcal{J}^a(x)\equiv\left[(\delta^{ab}+2G(\phi)^{ab})\frac{F\square\phi^b}{\sqrt{2}}-\partial\phi^T\cdot \frac{F \, \partial_\phi G(\phi)^a}{ \sqrt{2}}\cdot\partial\phi+2  \frac{F\,\partial G(\phi)}{ \sqrt{2}}\cdot\partial\phi^a\right]-\partial_\mu J^{\mu, \, a}_\text{pp} \ . \label{JderDip}
\end{align}
Now, we can read off the radiation field at $r\rightarrow\infty$ in terms of the Fourier transform of the source:
\begin{equation}
\phi^a=-\frac{\sqrt{2}}{F}\int\frac{\ud^d k}{(2\pi)^d}\, \frac{e^{-i\,k\cdot x}}{k^2} \mathcal{J}^a(k) \ .
\end{equation}
As before, the initial configuration consists of $N$ particles that are moving with constant velocity at $s=-\infty$, and the initial conditions for the color-charged point-particles are:
\begin{align}
x^\mu_\alpha|_{s\rightarrow-\infty}&=b^\mu_\alpha+p^\mu_\alpha s \ ,\\
{M^a_\alpha}^\mu|_{s\rightarrow-\infty}&={M^a_\alpha}^\mu \ .
\end{align}
From \eqref{JderDip} we can see that the $\mathcal{O}(1)$ field is
\begin{equation}
\phi^a\big{|}_{\mathcal{O}(1)}=i\frac{2}{F}\sum_{\alpha}m_\alpha^2\int_q\frac{e^{-iq\cdot(x-b_\alpha)}}{q^2}\,(q\cdot{M^a_\alpha})\,2\pi \delta(q\cdot p_\alpha)\ . \label{phi12Dip}
\end{equation}
Similarly to the previous case, we compute the deflections $\bar{x}_\alpha^\mu$ and $k\cdot \bar{M}_\alpha^a$ defined by
\begin{align}
x^\mu_\alpha(s)&=b^\mu_\alpha+p^\mu_\alpha s+ \bar{x}_\alpha^\mu(s) \ ,\\
k\cdot M_\alpha^a(s)&=k\cdot M_\alpha^a+k\cdot \bar{M}_\alpha^a(s)\ ,
\end{align}
where the barred quantities vanish at $s=-\infty$. At leading order, these are given by:
\begin{align}
\bar x^\mu_\alpha\big{|}_{\mathcal{O}(2)}&=-i\frac{4}{ F^2}\sum_{\beta\neq\alpha} \int\frac{\ud^d q}{(2\pi)^d}\,m_\alpha^2m_\beta^2 q^\mu (q\cdot M_\alpha^a) \,(q\cdot M_\beta^a)\,\frac{e^{-i\,q\cdot\left(b_{\alpha\beta}+p_\alpha s\right)}}{q^2 \, (q\cdot p_\alpha)^2} \, 2\pi \delta(q\cdot p_\beta) \ , \label{x2derDip} \\
(k\cdot \bar{M}_\alpha^a)\big{|}_{\mathcal{O}(2)}&=i\frac{2}{F^2} f^{abc} \sum_{\beta\neq\alpha} \int\frac{\ud^d q}{(2\pi)^d}\,m_\beta^2 \left[(k-2q)\cdot M_\alpha^b\right] \,(q\cdot M_\beta^c)\, \frac{e^{-i\,q\cdot\left(b_{\alpha\beta}+p_\alpha s\right)}}{q^2} \, 2\pi \delta(q\cdot p_\beta)  \ . \label{c2derDip}
\end{align}
Once we have the deflections at this order, we may compute $\mathcal{J}^a(k)$ at $\mathcal{O}(2)$:
\begin{align}
\mathcal{J}_\text{p.p.}^a(k) \big{|}_{\mathcal{O}(2)}=&\sqrt{2}\sum_{\alpha}\int \ud s\, e^{i\,k\cdot\left(b_\alpha+p_\alpha s \right)}m_\alpha^2\Big[k\cdot \bar{x}_\alpha(s)\big{|}_{\mathcal{O}(F'^{-2})}(k\cdot M_\alpha^a)-i(k\cdot \bar{M}_\alpha^a(s))\big{|}_{\mathcal{O}(F^{-2})}\nonumber \\
&-i\frac{1}{F}f^{abc}(k\cdot M_\alpha^b)\phi^c\big{|}_{\mathcal{O}(F'^{-1})}\Big] \ ,
\end{align}
which, after substituting in the corresponding deflections, becomes
\begin{equation}
\mathcal{J}_\text{p.p.}^a(k) \big{|}_{\mathcal{O}(2)}=i\sqrt{2}\frac{4}{F^2}\sum_{\alpha, \, \beta\neq\alpha}\int_{q_\alpha, \, q_\beta}\,m_\alpha^2m_\beta^2 \rho^a_{\alpha\beta}(k) \, \mu_{\alpha,\beta}(k) \ , \label{JNLSM2der}
\end{equation}
where
\begin{equation}
\rho^a_{\alpha\beta}(k)\equiv q_\alpha^2\Bigg[m_\alpha^2\,(k\cdot M_\alpha^a) (q_\beta\cdot M_\alpha^b) \,(q_\beta\cdot M_\beta^b) \,  \frac{k\cdot q_\beta}{(k\cdot p_\alpha)^2}+i \, f^{abc} (q_\alpha\cdot M_\alpha^b) (q_\beta\cdot M_\beta^c) \Bigg] \ .
\end{equation}
Once again, this expression is not manifestly invariant under permutations of the particle indices. Such invariance is ensured by the sum over the indices $\alpha$ and $\beta$.

We now proceed to compute the deflections of the point-particle coordinates and the dipole at next to leading order. These are given by 
\begin{align}
\bar x^\mu_\alpha\big{|}_{\mathcal{O}(4)}=&\frac{8}{ F^4} m_\alpha^2\underset{\gamma \neq\{\alpha,\beta\}}{\sum_{\beta \neq\alpha,}} \underset{q_\beta, \, q_\gamma\, q_\delta}{\int}\frac{ e^{-i\,q_\delta\cdot\left(b_{\alpha}+p_\alpha s\right)} }{\left( q_\delta\cdot p_\alpha\right)^2}\mu_{\beta,\gamma}(q_\delta)m_\beta^2 m_\gamma^2 \Biggl[i\,2\,(q_\delta\cdot M_\alpha^a) q_\delta^\mu \rho_{\beta\gamma}^a(q_\delta) \nonumber \\
&+f^{abc}((q_\gamma-2q_\beta)\cdot {M^b_\alpha})(q_\beta\cdot M_\beta^c)(q_\gamma\cdot M_\gamma^a) q_\gamma^\mu-f^{abc}(q_\beta\cdot {M^b_\beta})(q_\gamma\cdot M_\gamma^c)(q_\beta\cdot M_\alpha^a) q_\delta^\mu \nonumber \\
&+i\,2\,q_\gamma^\mu m_\alpha^2 \frac{q_\beta\cdot q_\alpha}{(q_\beta\cdot p_\alpha)^2}(q_\beta\cdot M_\alpha)\cdot(q_\beta\cdot M_\beta) \,\,(q_\gamma\cdot M_\alpha) \cdot(q_\gamma\cdot M_\gamma) \Biggl] \label{x4der} \\
k\cdot\bar{M}^a_\alpha\big{|}_{\mathcal{O}(4)}=&- i \frac{8}{F^4} f^{abc} \underset{\gamma \neq\{\alpha,\beta\}}{\sum_{\beta \neq\alpha,}} \underset{q_\beta, \, q_\gamma\, q_\delta}{\int} \frac{ e^{-i\,q_\delta\cdot\left(b_{\alpha}+p_\alpha s\right)} }{\left( q_\delta\cdot p_\alpha\right)}\mu_{\beta,\gamma}(q_\delta)m_\beta^2 m_\gamma^2\Biggl[\frac{1}{q_\delta^2}((k-2q_\delta\cdot M^b))\rho_{\beta\gamma}^c(q_\delta) \nonumber \\
&+m_\alpha^2 \,((k-2q_\gamma)\cdot M^b)(q_\gamma\cdot M_\gamma^c)\, (q_\beta\cdot M_\alpha)\cdot(q_\beta\cdot M_\beta)\frac{q_\beta \cdot q_\gamma}{(q_\beta\cdot p_\alpha)^2} \nonumber \\
&-if^{bde}((\frac{7}{6}k-q_\beta)\cdot {M^d_\alpha})(q_\beta\cdot M_\beta^e)(q_\gamma\cdot M_\gamma^c)  \Biggl] \ . \label{c4der}\end{align}
\begin{figure}[!t]
	\includegraphics[scale=0.9]{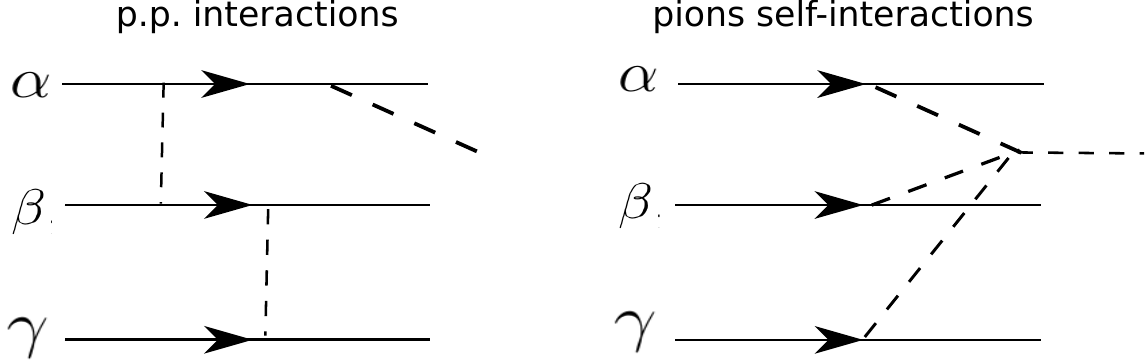}
	\caption{Interactions contributing to the NLSM source at $\mathcal{O}(4)$. The left-hand side shows the contribution from the interactions between the point-particles and pions and the right-hand side shows the contribution from the pion 4-point vertex.}
	\label{NLSMorder4}
\end{figure}
Once we have the deflections at next to leading order, we can use them to compute the current at $\mathcal{O}(4)$. There are two contributions to this current, one coming from self-interactions of the NLSM and the other one coming from the coupling to the point-particles. The contribution to the current from the interactions with the point-particles reads
\begin{align}
\mathcal{J}_{p.p.}^a\big{|}_{\mathcal{O}(4)}(k)=&-i\sqrt{2}\int \ud s\, e^{i\,k\cdot\left(b_\alpha+p_\alpha s \right)} m_\alpha^2\Bigg[i \,k\cdot \bar{x}_\alpha(s)\big{|}_{\mathcal{O}(4)}(k\cdot M_\alpha^a)+(k\cdot \bar{M}_\alpha^a(s))\big{|}_{\mathcal{O}(4)}\nonumber \\
&+i k\cdot \bar{x}_\alpha(s)\big{|}_{\mathcal{O}(2)}(k\cdot \bar{M}_\alpha^a)\big{|}_{\mathcal{O}(2)}+\frac{1}{2}(ik\cdot \bar{x}_\alpha(s)\big{|}_{\mathcal{O}(2)})^2(k\cdot M_\alpha^a) \nonumber  \\
&+\frac{1}{F}f^{abc}(k\cdot\bar{M}_\alpha^b)\big{|}_{\mathcal{O}(2)}\phi^c\big{|}_{\mathcal{O}(1)} +\frac{2}{3F^2}f^{abc}f^{bde}(k\cdot M_\alpha^d)\phi^e\big{|}_{\mathcal{O}(1)}\phi^c\big{|}_{\mathcal{O}(1)} \nonumber \\
&+\frac{1}{F}f^{abc}(k\cdot M_\alpha^b)\left(i k\cdot \bar{x}_\alpha(s)\big{|}_{\mathcal{O}(2)}\phi^c\big{|}_{\mathcal{O}(1)}+\phi^c\big{|}_{\mathcal{O}(3)}\right)\Bigg]  \ , \label{J4NLSMder}
\end{align} 
and corresponds to the left-hand side diagram of Fig. \ref{NLSMorder4}. Meanwhile, the contribution from the NLSM self-interactions, which comes from the right-hand side diagram in Fig. \ref{NLSMorder4}, is 
\begin{equation}
\mathcal{J}_\text{s.i.}^a(x) \big{|}_{\mathcal{O}(4)}=\frac{\sqrt{2}}{ 3\, F}f^{abc}f^{cde}\phi^e\big{|}_{\mathcal{O}(1)}\,\left(\partial^\mu\phi^b\big{|}_{\mathcal{O}(1)}\, \partial_\mu\phi^d\big{|}_{\mathcal{O}(1)} +\frac{1}{2}\square\phi^d\big{|}_{\mathcal{O}(1)}\, \phi^b\big{|}_{\mathcal{O}(1)} \right) \ .
\end{equation}
After using the leading order contribution to the NLSM field we find 
\begin{align}
\mathcal{J}_\text{s.i.}^a(k) \big{|}_{\mathcal{O}(4)}=-i\frac{8\sqrt{2}}{  3\,F^4}\underset{\gamma \neq\{\alpha,\beta\}}{\sum_{\alpha,\,\beta \neq\alpha,}} \underset{q_\alpha, \, q_\beta, \, q_\gamma\, q_\delta}{\int}(2\pi)^d\delta^d(q_\delta-q_\beta-q_\gamma)\,\mu_{\alpha,\beta,\gamma}(k)m_\alpha^2m_\beta^2m_\gamma^2\sigma^a_{\alpha\beta\gamma} \ ,
\end{align}
where
\begin{equation}
\sigma^a_{\alpha\beta\gamma}=f^{abc}f^{cde}(q_\alpha\cdot{M^b_\alpha}) (q_\beta\cdot{M^d_\beta}) (q_\gamma\cdot{M^e_\gamma})\left(q_\alpha\cdot q_\beta+\frac{1}{2}q_\beta^2\right) \ .
\end{equation} 
Notice that if we set ${M^a}^\mu=c^ap^\mu$ then there is no radiation at any order. This is understood by realizing that the coupling $c_a p^\mu\nabla_\mu\phi^a$ arises from the Yang-Mills gauge invariant coupling $c^ap^\mu A^a_\mu$ after introducing the St\"{u}ckelberg field. Since the starting point is a gauge invariant term, when we introduce the St\"{u}ckelberg field we are simply performing a gauge transformation and thus the physics does not change.

\section{Single copy: bi-adjoint scalar to non-linear sigma model} \label{singlecopy}
So far, we have computed the radiation amplitude at order $\mathcal{O}(4)$ for both the bi-adjoint scalar and the non-linear sigma model. It is now possible to identify the generalized color-kinematics replacements needed to obtain the single copy. Since the color degrees of freedom of the group $G$ are different in the bi-adjoint scalar and the NLSM, we need to perform replacements to take the color charges to the color dipoles. Schematically, the color-kinematics and color-color replacements that we use are
\begin{equation}
\tilde{C}(\tilde{c}^{\tilde{a}})\rightarrow \tilde{N}(\{q\}) \ ,  \quad C(\{q\};c^a)\rightarrow C(\{q\}\cdot M^a).
\end{equation}
From dimensional analysis we know that when transforming from the color charges to the color dipoles for the radiation at $\mathcal{O}(3)$ we need to have a factor with mass dimension six on the NLSM side. This is due to the discrepancy in the mass dimension of the couplings of these theories. Considering this, we find that the replacements are:
\begin{subequations}
	\begin{align}
	c_\alpha\,( c_\alpha\cdot c_\beta)&\rightarrow -i \,2 \sqrt{2}\,m_\alpha^4 m_\beta^2 \,(k\cdot M_\alpha)(q_\beta\cdot M_\alpha)\cdot(q_\beta\cdot M_\beta) \ ,\\
	\frac{f^{abc}c_\alpha^bc_\beta^c}{q_\beta\cdot p_\alpha}&\rightarrow -i \,2 \sqrt{2}\,m_\alpha^2 m_\beta^2 \, \,f^{abc}(q_\alpha\cdot M_\alpha)^b(q_\beta\cdot M_\beta)^c \ ,
	\end{align}
	\label{CC2}
\end{subequations}
and
\begin{subequations}
	\begin{align}
	\tilde{c}_\alpha \,( \tilde{c}_\alpha\cdot \tilde{c}_{\beta})\rightarrow 1 \ , \\
	\tilde{f}^{\tilde{a}\tilde{b}\tilde{c}}\tilde{c}_\alpha^{\tilde{b}} \tilde{c}_\beta^{\tilde{c}}\rightarrow 0 \ .
	\end{align}
	\label{CK2}
\end{subequations}
The fact that the terms involving one structure constant are set to zero can be traced back to the fact that the non-linear sigma model cubic vertex is zero.  Under these replacements, we can see that on-shell:
\begin{equation}
\rho_{\alpha\beta}^{a\,\tilde{a}}(k)\rightarrow i \,2 \sqrt{2}\,m_\alpha^2m_\beta^2\, \rho_{\alpha\beta}^{a}(k) \ ,
\end{equation}
which implies that the radiation amplitude for the bi-adjoint scalar at $\mathcal{O}(2)$, Eq.\eqref{J2BS}, maps to the radiation amplitude of the NLSM at $\mathcal{O}(2)$, Eq.\eqref{JNLSM2der}.

In order to find the mapping for the radiation at order $\mathcal{O}(5)$ we require color-kinematics and color-color replacements for color factors involving contractions of five color structures. In this case, the NLSM side should have mass dimension ten. These replacements are given by
\begin{subequations}
	\footnotesize{
		\begin{align}
		c_\alpha^a (c_\alpha\cdot c_\beta)(c_\alpha\cdot c_\gamma)&\rightarrow i \,4\,\sqrt{2}\, m_\alpha^6m_\beta^2m_\gamma^2 (k\cdot M_\alpha)^a(q_\beta\cdot M_\alpha)\cdot(q_\beta\cdot M_\beta) (q_\gamma\cdot M_\alpha) \cdot(q_\gamma\cdot M_\gamma)  \ ,  \\
		c_\alpha^a (c_\alpha\cdot c_\beta)(c_\beta\cdot c_\gamma)&\rightarrow i \,4\,\sqrt{2}\, m_\alpha^4m_\beta^4m_\gamma^2 (k\cdot M_\alpha)^a(q_\delta\cdot M_\alpha)\cdot(q_\delta\cdot M_\beta) (q_\gamma\cdot M_\beta) \cdot(q_\gamma\cdot M_\gamma)   \ , \\
		\frac{c_\alpha^a\,\,\,( f\cdot c_\alpha\cdot c_\beta \cdot c_\gamma)}{q_\beta\cdot p_\alpha}&\rightarrow i \,4\,\sqrt{2}\, m_\alpha^4 m_\beta^2m_\gamma^2 (k\cdot M_\alpha) (f\cdot((q_\gamma\!-\!3q_\beta/2)\cdot M_\alpha)\cdot(q_\beta\cdot M_\beta)\cdot (q_\gamma\cdot M_\gamma) ) \ ,  \\
		\frac{c_\alpha^a
			\,\,\,( f\cdot c_\alpha\cdot c_\beta \cdot c_\gamma)}{q_\beta\cdot p_\gamma}&\rightarrow i \,4\,\sqrt{2}\, m_\alpha^4 m_\beta^2m_\gamma^2 (k\cdot M_\alpha) (f\cdot(q_\delta\cdot M_\alpha)\cdot(q_\beta\cdot M_\beta)\cdot ((q_\beta-q_\delta)\cdot M_\gamma) )  \ , \\
		\frac{(c_\alpha\cdot c_\gamma) f^{abc}  c_\alpha^b  c_\beta^c }{q_\beta\cdot p_\alpha}&\rightarrow i \,4\,\sqrt{2}\, m_\alpha^4 m_\beta^2m_\gamma^2 (q_\gamma\cdot M_\alpha)\cdot (q_\gamma\cdot M_\gamma)f^{abc}((k-q_\beta)\cdot M_\alpha)^b(q_\beta\cdot M_\beta)^c )  \ , \\
		\frac{(c_\alpha\cdot c_\gamma) f^{abc}  c_\alpha^b  c_\beta^c }{q_\delta\cdot p_\alpha}&\rightarrow i \,4\,\sqrt{2}\, m_\alpha^4 m_\beta^2m_\gamma^2 (q_\gamma\cdot M_\alpha)\cdot (q_\gamma\cdot M_\gamma) f^{abc}((k-q_\beta)\cdot M_\alpha)^b(q_\beta\cdot M_\beta)^c )  \ , \\
		\frac{(c_\alpha\cdot c_\gamma) f^{abc}  c_\alpha^b  c_\beta^c }{q_\delta\cdot p_\beta}&\rightarrow i \,4\,\sqrt{2}\, m_\alpha^4 m_\beta^2m_\gamma^2 (q_\gamma\cdot M_\alpha)\cdot (q_\gamma\cdot M_\gamma)\, f^{abc}(q_\delta\cdot M_\alpha)^b((k-q_\delta)\cdot M_\beta)^c )  \ , \\
		\frac{f^{abc}f^{bde}c_\alpha^dc_\beta^ec_\gamma^c}{(q_\beta\cdot p_\alpha)(q_\delta\cdot p_\alpha)}&\rightarrow i \,4\,\sqrt{2}\, m_\alpha^2 m_\beta^2m_\gamma^2 f^{abc}f^{bde}\left[(k-\frac{q_\delta}{2}+q_\alpha\frac{1}{12 q_\gamma^2} n(\alpha,\beta,\gamma)\right]\cdot M_\alpha)^d(q_\beta\cdot M_\beta)^e (q_\gamma\cdot M_\gamma)^c \ , \\
		\frac{f^{abc}f^{bde}c_\beta^dc_\gamma^ec_\alpha^c}{(q_\beta\cdot p_\gamma)(q_\delta\cdot p_\alpha)}&\rightarrow i \,4\,\sqrt{2}\, m_\alpha^2 m_\beta^2m_\gamma^2 f^{abc}f^{bde} (q_\beta\cdot M_\beta)^d (q_\gamma\cdot M_\gamma)^e\left[\left(k-q_\delta+q_\alpha\frac{q_\delta^2}{12 q_\alpha^2q_\gamma^2} n(\alpha,\beta,\gamma) \right)\cdot M_\alpha\right]^c \ ,
		\end{align}
	}
\label{CC4}
\normalsize{}
\end{subequations}
where $q_\delta=q_\beta+q_\gamma$, $n(\alpha,\beta,\gamma)$ is the 4-point amplitude of the non-linear sigma model
\begin{equation}
n(\alpha,\beta,\gamma)\equiv \frac{(q_\beta+q_\alpha)^2-(q_\gamma+q_\alpha)^2}{3} \ ,
\end{equation}
 and
\begin{subequations}
	\begin{align}
	\tilde{c}_\alpha^{\tilde{a}} (\tilde{c}_\alpha\cdot \tilde{c}_\beta)(\tilde{c}_\alpha\cdot \tilde{c}_\gamma)&\rightarrow 1 \ , \\
	\tilde{c}_\alpha^{\tilde{a}} (\tilde{c}_\alpha\cdot \tilde{c}_\beta)(\tilde{c}_\beta\cdot \tilde{c}_\gamma)&\rightarrow 1 \ , \\
	\tilde{c}_\alpha^{\tilde{a}} \,\,\,( \tilde{f}\cdot \tilde{c}_\alpha\cdot \tilde{c}_\beta \cdot \tilde{c}_\gamma)&\rightarrow 0 \ , \\
	(\tilde{c}_\alpha\cdot \tilde{c}_\gamma) \tilde{f}^{\tilde{a}\tilde{b}\tilde{c}}  \tilde{c}_\alpha^{\tilde{b}}  \tilde{c}_\beta^{\tilde{c}}&\rightarrow 0 \ , \\
	\tilde{f}^{\tilde{a}\tilde{b}\tilde{c}}f^{\tilde{b}\tilde{d}\tilde{e}}\tilde{c}_\beta^{\tilde{d}}\tilde{c}_\gamma^{\tilde{e}}\tilde{c}_\alpha^{\tilde{c}}&\rightarrow 0\ .
	\end{align}
	\label{CK4}
\end{subequations}
Under these, the on-shell current of Eq.\eqref{J4BS} is mapped to the on-shell current of Eq.\eqref{J4NLSMder}. Note that, in the color-color replacements we see that the left side involves denominators when the structure constants are present. These factors appeared due to the difference between the equations of motion for the color charges, where an integration gives rise to denominators, and the dipoles, which are contracted with momentum factors. There is also an ambiguity in how to pick the splitting in the replacement rules since now we also map the color factors to new color factors. By shifting kinematic factors from the color-color to the color-kinematics replacements it is possible to find a different set of rules that give the desired map between theories.

\section{Special Galileon radiation} \label{SG}

In this section, we compute the scalar radiation generated by point-particles coming from infinity that are coupled to the special Galileon. The Lagrangian for the special Galileon theory is~\cite{Hinterbichler:2015pqa}
\begin{equation}
\lag_\text{SG}=\frac{1}{2}\left(\partial\pi\right)^2-\frac{1}{12\Lambda^{6}}\left(\partial\pi\right)^2\left[\left(\square\pi\right)^2-\left(\partial_\mu\partial_\nu \pi\right)^2\right]+\cdots \ ,
\end{equation}
where $\Lambda$ is the strong coupling scale. In four dimensions it only contains the quartic Galileon term but it includes higher order terms in higher dimensions. The action for the special Galileon is invariant under
\begin{align}
\delta\pi&=c+b_\mu x^\mu + s_{\mu\nu}x^\mu x^\nu+\frac{1}{\Lambda^{6}}s^{\mu\nu}\partial_\mu\pi \partial_\nu\pi \ ,
\end{align}
where $c$ is a constant, $b_\mu$ is a constant vector, and $s_{\mu\nu}$ is a traceless symmetric constant tensor. As we mentioned previously, we assume that the special Galileon is coupled to the point-particles through a conformal rescaling of the metric $g_{\mu\nu}\rightarrow(1+2\pi/\Lambda)g_{\mu\nu}$. This is motivated by the coupling that arises in the decoupling limit of massive gravity for Galileons. Hence, the point-particle action is:
\begin{equation}
S_\text{pp}=-\sum_{\alpha}m_\alpha\int \ud \lambda\sqrt{ 1+2\frac{\pi}{\Lambda}} \sqrt{\frac{\ud x_\alpha}{\ud \lambda}\cdot\frac{\ud x_\alpha}{\ud \lambda}} \ , \label{ppgal}
\end{equation}
or, by introducing the einbein $\eta$,
\begin{equation}
S_\text{pp}=-\frac{1}{2}\sum_{\alpha}\int \ud \lambda \left(\eta^{-1}(\lambda)\,\frac{\ud x_\alpha}{\ud \lambda}\cdot\frac{\ud x_\alpha}{\ud \lambda}+\eta(\lambda)m_\alpha^2\right)\sqrt{1+2\frac{\pi}{\Lambda}} \ . 
\end{equation}
If we assumed that this interaction arises from the decoupling limit of massive gravity, then the coupling strength would be $1/M_{Pl}$. In this case, the couplings of the Galileon with itself and with the point-particles would be suppressed by two different scales, satisfying the hierarchy $\Lambda\ll M_{Pl}$.  The current calculation for the radiation amplitude will not hold in this case since the leading terms would come from the self-interactions. We find that in order to identify the special Galileon as the double copy of the NLSM, the coupling with the point-particles should have the same strength as the self-interactions coupling therefore we assume the interactions in Eq.\ref{ppgal}. The equation of motion that determines the evolution of the point-particle coordinates $x^\mu$ is
\begin{equation}
\frac{\ud p^\mu_\alpha}{\ud s}-\frac{\partial_\nu\pi}{\Lambda \left(1+2\frac{\pi}{\Lambda^{d/2-1}}\right)}\left(m_\alpha^2 \, \delta^{\mu\nu} -p^\mu_\alpha p_\alpha^\nu\right)=0 \  , \label{ppSG}
\end{equation}
while the equation of motion for the special Galileon is 
\begin{align}
\square\pi&-\frac{1}{6 \Lambda^{6}}\left[(\square\pi)^3+2\left(\partial_\mu\partial_\nu\pi\right)^3-3(\square\pi)\left(\partial_\mu\partial_\nu\pi\right)^2\right] +\frac{1}{\Lambda}\sum_{\alpha}\int\ud s \, \tfrac{m_\alpha^2}{\sqrt{1+2\frac{\pi}{\Lambda}}}  \delta^d(x-x_\alpha(s))=0 \ .
\end{align}

\subsection{Perturbative solutions}
We compute the solution in powers of the coupling constant, but the actual perturbation parameter for the special Galileon is 
\begin{equation}
\epsilon\propto \frac{1}{\Lambda^{2}}\frac{E}{b}
\end{equation}
Rewriting the special Galileon equation of motion as
\begin{equation}
\square\pi=\frac{1}{\Lambda}\mathcal{J} \ ,
\end{equation}
where the source current $\mathcal{J}$ is defined as
\begin{equation}
\mathcal{J}\equiv\frac{1}{6 \Lambda^5}\left[(\square\pi)^3+2\left(\partial_\mu\partial_\nu\pi\right)^3-3(\square\pi)\left(\partial_\mu\partial_\nu\pi\right)^2\right]-\sum_{\alpha}\int\ud s \,\frac{m_\alpha^2}{\sqrt{1+\frac{2\pi}{\Lambda}}} \, \delta^d(x-x_\alpha(s)) \ ,
\end{equation}
we can write the leading order special Galileon field as
\begin{equation}
\pi (x) =-\frac{1}{\Lambda}\int\frac{\ud^d k}{(2\pi)^d}\, \frac{e^{-i\,k\cdot x}}{k^2} \mathcal{J}(k) \ .
\end{equation}
Note that the contributions from the special Galileon self-interactions will only appear at $\mathcal{O}(8)$. As before, the point-particles  are moving with constant velocity at $s=-\infty$ and thus the initial conditions are:
\begin{align}
x^\mu_\alpha|_{s\rightarrow-\infty}&=b^\mu_\alpha+p^\mu_\alpha s \ .
\end{align}
From this we find that the field at $\mathcal{O}(\Lambda^{-1})$ is given by
\begin{equation}
\pi\big{|}_{\mathcal{O}(1)}=\frac{1}{\Lambda}\sum_{\alpha}\int\frac{\ud^d k}{(2\pi)^d}\, m_\alpha^2 \frac{e^{-i\,k\cdot\left(x-b_\alpha\right)}}{k^2} \, 2\pi \delta(k\cdot p_\alpha)\ \ . \label{pi1}
\end{equation}
As in the previous cases, to obtain the $\mathcal{O}(\Lambda^{-2})$ field, we need to find the deflection
\begin{align}
x^\mu_\alpha&=b^\mu_\alpha+p^\mu_\alpha s+ \bar{x}_\alpha^\mu(s) \ .
\end{align} 
Using the $\mathcal{O}(\Lambda^{-1})$ field in the equation of motion \eqref{ppSG}, we find
\begin{align}
\bar x^\mu_\alpha\big{|}_{\mathcal{O}(2)}&=i\frac{1}{ \Lambda^2}\sum_{\beta\neq\alpha} m_\beta^2 \int\frac{\ud^d q}{(2\pi)^d}\, \left(m_\alpha^2\,q^\mu-(q\cdot p_\alpha)p_\alpha^\mu\right)\frac{e^{-i\,q\cdot\left(b_{\alpha\beta}+p_\alpha s\right)}}{q^2 \, (q\cdot p_\alpha)^2} \, 2\pi \delta(q\cdot p_\beta) \ , \label{x2SG}
\end{align}
which leads to the current:
\begin{align}
&\mathcal{J}(k) \big{|}_{\mathcal{O}(2)}=\frac{1}{\Lambda^2}\sum_{\alpha, \, \beta\neq\alpha}\int_{q_\alpha, \, q_\beta}\, m_\alpha^2 m_\beta^2 \, \mu_{\alpha,\beta}(k)  \rho_{\alpha\beta}(k) \ , \label{J2SG}
\end{align}
where
\begin{equation}
\rho_{\alpha\beta}(k) \equiv -m_\alpha^2q_\alpha^2\frac{k\cdot q_\beta}{(k\cdot p_\alpha)^2} \ .
\end{equation}
This current comes from the point-particle interactions diagram in Fig.\ref{order2}. We can now obtain the $\mathcal{O}(4)$ current which comes from the point-particle interactions diagram in Fig.\ref{NLSMorder4} and is given by 
\begin{align}
\mathcal{J}\big{|}_{\mathcal{O}(4)}=&\sum_{\alpha}m_\alpha^2\int \ud s\, e^{i\,k\cdot\left(b_\alpha+p_\alpha s \right)}\Big[-i k\cdot \bar{x}_\alpha\big{|}_{\mathcal{O}(4)}+\Lambda^{-1}\pi\big{|}_{\mathcal{O}(3)}+\Lambda^{-1}\bar{x}^\mu_\alpha\big{|}_{\mathcal{O}(2)}\partial_\mu\pi\big{|}_{\mathcal{O}(1)} \nonumber \\
&+i\Lambda^{-1} k\cdot \bar{x}_\alpha(s)\big{|}_{\mathcal{O}(2)}\pi\big{|}_{\mathcal{O}(1)}-\frac{3}{2}\Lambda^{-2}{\pi\big{|}_{\mathcal{O}(1)}^2}+\frac{1}{2}\left(-i k\cdot \bar{x}_\alpha\big{|}_{\mathcal{O}(2)}\right)^2\Big] \ ,  \label{J4SG}
\end{align}
where
\begin{align}
\bar x^\mu_\alpha\big{|}_{\mathcal{O}(4)}=&\frac{i}{\Lambda^4} \underset{\gamma \neq\{\alpha,\beta\}}{\sum_{\beta \neq\alpha,}} \underset{q_\beta, \, q_\gamma\, q_\delta}{\int}\frac{ e^{-i\,q_\delta\cdot\left(b_{\alpha}+p_\alpha s\right)} }{\left( q_\delta\cdot p_\alpha\right)^2}m_\beta^2 m_\gamma^2\,\,\mu_{\beta,\gamma}(q_\delta)\Biggl[ \frac{1}{q_\gamma\cdot p_\alpha}\Big(-m_\alpha^2q_\beta^\nu\left(q_\gamma^\nu p_\alpha^\mu+ p_\alpha^\nu q_\gamma^\mu\right) \nonumber \\
 &+2(q_\beta\cdot p_\alpha)(q_\gamma\cdot p_\alpha) p_\alpha^\mu\Big)+\left(-m_\alpha^2\delta^\mu_\nu+p_\alpha^\mu {p_\alpha}_\nu\right)\Bigg(\frac{1}{q_\delta^2}\rho_{\beta\gamma}(q_\delta)q_\delta^\nu \nonumber \\
 &+2q_\gamma^\nu+\frac{1}{(q_\gamma\cdot p_\alpha)^2}\left(-m_\alpha^2 q_\beta\cdot q_\gamma+(q_\beta\cdot p_\alpha)(q_\gamma \cdot p_\alpha)\right)q_\gamma^\nu\Bigg)\Biggl] \ .
\end{align}
Once we substitute the corresponding trajectory deviations and field profiles into the equation for the current at next to leading order, we find that the terms whose mass factors are of the form $m^2_\alpha m^2_\beta m^2_\gamma$ and $m^4_\alpha m^2_\beta m^2_\gamma$   become zero after using the corresponding delta functions.

\section{Double copy: non-linear sigma model to special Galileon} \label{doublecopy}
In the previous section we computed the special Galileon radiation at next to leading order and previously we performed the analogue calculation for the NLSM. We are now in position to propose the color-kinematics replacements that lead to the double copy. In this case, the replacements are of the form
\begin{equation}
C(\{q\}\cdot M^a)\rightarrow N(\{q\}) \ .
\end{equation}
Specifically, to transform the NLSM to the special Galileon at order $\mathcal{O}(2)$  we use the color-kinematics replacements:
\begin{subequations}
	 \begin{align}
	-i \,2 \sqrt{2}\,(k\cdot M_\alpha)(q_\beta\cdot M_\alpha)\cdot(q_\beta\cdot M_\beta) &\rightarrow 1\ ,\\
	-i \,2 \sqrt{2}\,f^{abc}(q_\alpha\cdot M_\alpha)^b(q_\beta\cdot M_\beta)^c&\rightarrow 0 \ ,
	\end{align}
	\label{DCK2}
\end{subequations}
Under these replacements, we can see that on-shell:
\begin{equation}
 \rho_{\alpha\beta}^{a}(k)\rightarrow\rho_{\alpha\beta}(k) \ ,
\end{equation}
and thus the radiation from Eq.\eqref{JNLSM2der} is mapped to that of Eq.\eqref{J2SG}. At $\mathcal{O}(4)$ we need replacements for higher order contractions of dipole moments. These color-kinematics replacements are:
\small{
\begin{subequations}
	\begin{align}
	i \,4\,\sqrt{2}\, &(k\cdot M_\alpha)^a(q_\beta\cdot M_\alpha)\cdot(q_\beta\cdot M_\beta) (q_\gamma\cdot M_\alpha) \cdot(q_\gamma\cdot M_\gamma) \rightarrow  -1 \ ,  \\
	i \,4\,\sqrt{2}\,  &(k\cdot M_\alpha)^a(q_\delta\cdot M_\alpha)\cdot(q_\delta\cdot M_\beta) (q_\gamma\cdot M_\beta) \cdot(q_\gamma\cdot M_\gamma) \rightarrow -1  \ , \\
	i \,4\,\sqrt{2}\, & (k\cdot M_\alpha) [f\cdot(q_\gamma\cdot M_\alpha)\cdot(q_\beta\cdot M_\beta)\cdot (q_\gamma\cdot M_\gamma) ]\rightarrow  0\ ,  \\
	i \,4\,\sqrt{2}\,& (k\cdot M_\alpha) [f\cdot(q_\delta\cdot M_\alpha)\cdot(q_\beta\cdot M_\beta)\cdot (q_\gamma\cdot M_\gamma) ] \rightarrow0  \ , \\
	i \,4\,\sqrt{2}\,&(q_\gamma\cdot M_\alpha)\cdot (q_\gamma\cdot M_\gamma)f^{abc}(q\cdot M_\alpha)^b(q_\beta\cdot M_\beta)^c \rightarrow  0 \ , \\
	i \,4\,\sqrt{2}\,& (q_\gamma\cdot M_\alpha)\cdot (q_\gamma\cdot M_\gamma)\, f^{abc}(q\cdot M_\beta)^b(q_\delta\cdot M_\alpha)^c \rightarrow 0 \ , \\
	i \,4\,\sqrt{2}\,  &f^{abc}f^{bde}(q_\delta\cdot M_\alpha)^d(q_\beta\cdot M_\beta)^e (q_\gamma\cdot M_\gamma)^c \rightarrow -2\, n(\gamma,\alpha,\beta)\left(1+\frac{ q_\beta q_\gamma^2}{ q_\delta q_\alpha^2}+\frac{n(\gamma,\alpha,\beta)}{8 q_\gamma^2}\right)\ , \\
	i \,4\sqrt{2}\,  &f^{abc}f^{bde} (q_\beta\cdot M_\beta)^d (q_\gamma\cdot M_\gamma)^e(q_\alpha\cdot M_\alpha)^c \rightarrow n(\alpha,\beta,\gamma) \ , \label{Jacobi}
	\end{align}
	\label{DCK4}
\end{subequations}
}
\normalsize{}
where $q$ represents either the radiated momentum $k$, or the momenta any of the fields, such as $q_\alpha$. Under these replacements the NLSM radiation from Eq. \eqref{J4NLSMder} is transformed into the special Galileon one from Eq. \eqref{J4SG}. While the color factors involving only one structure constant are set to zero as a consequence of the lack of a cubic interaction in both the NLSM and the special Galileon, the color factors involving two structure constants give rise to a more interesting relation. The replacement in Eq. \eqref{Jacobi} exchanges the color factor of the four-point amplitude of the NLSM (from the right-hand side diagram in Fig. \ref{NLSMorder4})  with the corresponding color-stripped amplitude. Note that both the color and kinematic sides satisfy the Jacobi identity even though none of the involved momenta $\{q\}$ are on shell.

\section{Discussion and outlook} \label{conclusions}

In this paper we have computed the amplitude of radiation emitted by point-like particles coupled to a bi-adjoint scalar, a set of pions, and a special Galileon at next-to-leading order in the couplings. While one might naively expect that the NLSM coupling to point-particles should be of the form $c^a p^\mu \partial_\mu \phi^a$, this coupling gives rise to no radiation at any perturbative order. This can be understood from the fact that this coupling arises from the Yang-Mills gauge invariant coupling $c^ap^\mu A_\mu^a$ after introducing the St\"{u}ckelberg field $\phi$. A similar situation happens for the special Galileon and the coupling $p^\mu p^\nu \partial_\mu\partial_\nu\pi$. Instead, for the NLSM we consider a coupling to a color dipole moment, which is invariant under $G$, and for the special Galileon we consider the coupling arising after the conformal transformation $g_{\mu\nu}\rightarrow(1+2\pi/\Lambda)g_{\mu\nu}$. These couplings are motivated by those that would arise for longitudinal modes of massive Yang-Mills and massive gravity.  Using these couplings we have shown that up to next to leading order the double copy of the radiation of pions corresponds to the radiation of special Galileons. We have also constructed the single copy starting from the bi-adjoint scalar. To do so, we have used a generalized set of color-kinematics replacements to map from color charges to color dipole moments. Compared to the gravitational double copy, in the case of the scalar modes the simple replacement of the color structure by its corresponding vertex arises for the four-point case, since there are no cubic interactions.

At the order to which we have worked, the special Galileon self-interactions do not contribute to the radiation. It would be interesting to investigate the structure of the color-kinematics replacement rules at $\mathcal{O}(8)$, which is where the first contribution from these interactions is bound to appear. Our expectation is that at $\mathcal{O}(6)$ the color structures involving three structure constants will be set to zero, since these are related to the five-point vertex. At even higher order, the color structures with four structure constants---corresponding to the six-point vertex---will have a more complicated replacement rule, similar to what happens in the gravitational case at $\mathcal{O}(4)$.

At higher order one must also contend with another challenge, namely the fact that higher derivative corrections to the NLSM and special Galileon actions can contribute to the radiation field. This challenge is not limited to the classical double copy, since it is far from understood what are the correct higher derivative corrections that give rise to the amplitudes double copy relation~\cite{Elvang:2018dco,Padilla:2016mno,Rodina:2018pcb,Carrasco:2016ldy}. Another  question that would be interesting to explore is whether it is possible to construct a wider web of classical color-kinematic relations, in the spirit of \cite{Cachazo:2014xea,Cheung:2017ems}. This web should include not only Yang-Mills, gravity, and the scalar theories we have considered in this paper, but also other theories whose amplitudes admit a CHY representation---{\it e.g.} Born-Infeld theory, Dirac-Born-Infeld theory, and others. We plan to tackle some of these interesting questions in the near future.

\section{Acknowledgments}
We thanks Clifford Cheung and Walter Goldberger for useful discussions. This work was supported in part by US Department of Energy (HEP) Award DE-SC0013528. MT was also supported in part by NASA ATP grant NNH17ZDA001N.

\bibliographystyle{apsrev4-1}
\bibliography{bibliography}

\end{document}